\documentclass[sigconf]{acmart}

\copyrightyear{2021} 
\acmYear{2021} 
\setcopyright{acmlicensed}\acmConference[SIGMOD '21]{Proceedings of the 2021 International Conference on Management of Data}{June 20--25, 2021}{Virtual Event, China}
\acmBooktitle{Proceedings of the 2021 International Conference on Management of Data (SIGMOD '21), June 20--25, 2021, Virtual Event, China}
\acmPrice{15.00}
\acmDOI{10.1145/3448016.3457250}
\acmISBN{978-1-4503-8343-1/21/06}

\renewcommand\footnotetextcopyrightpermission[1]{}

\usepackage{balance}
\usepackage{flushend}
\usepackage{expl3}
\usepackage{amsmath}
\usepackage{url}
\usepackage{makecell}
\usepackage{scalerel}
\usepackage{epstopdf}
\usepackage{graphicx}
\usepackage{subfigure}
\usepackage{color}
\usepackage{tabularx}
\usepackage{hyperref}
\usepackage{ragged2e}  
\newcolumntype{Y}{>{\RaggedRight\arraybackslash}X}
\usepackage{booktabs}
\usepackage[noend]{algpseudocode}
\usepackage{fixltx2e}
\usepackage{paralist}
\usepackage{listings}
\usepackage{multirow}
\usepackage{enumerate}
\usepackage{bbm}
\usepackage{xspace}
\usepackage{comment}
\usepackage[inline]{aplcomments}
\usepackage{caption} 
\usepackage{tcolorbox} 
\usepackage[ruled,vlined]{algorithm2e}

\graphicspath{{../}}

\usepackage{comment}
\usepackage[labelfont=bf]{caption}
\MakeRobust{\Call}
\usepackage{enumitem}
\usepackage{etoolbox} 
\usepackage{mdwlist} 

\newenvironment{proofsketch}{\par{\noindent \textit{Proof Sketch:}}}{\qed\par}

\newcommand{\val}[1]{{\small \texttt{#1}}}

\newcommand{\sj}{\textsc{Auto-Validate}\xspace}

\newcommand{\rev}[1]{{\color{blue}#1}}

\DeclareMathOperator*{\avg}{avg}

\newcounter{definition}
\newenvironment{definition}[1][]{\refstepcounter{definition}\par\smallskip\textsc{Definition~\thedefinition.\ #1}}{\smallskip}

\newcounter{example}
\newenvironment{example}[1][]{\refstepcounter{example}\par\smallskip\textsc{Example~\theexample.\ #1}}{\smallskip}

\newcounter{theorem}
\newenvironment{theorem}[1][]{\refstepcounter{theorem}\par\smallskip\textsc{Theorem~\thetheorem.\ #1}}{\smallskip}

\newcounter{proposition}

\newtoggle{fullversion}
\toggletrue{fullversion}

\makeatletter
  \newcommand\figcaption{\def\@captype{figure}\caption}
  \newcommand\tabcaption{\def\@captype{table}\caption}
\makeatother

\begin{document}

\iftoggle{fullversion}{}{
\setlength{\floatsep}{0pt}
\setlength{\textfloatsep}{0pt}
\setlength{\abovecaptionskip}{0pt}
\setlength{\abovedisplayskip}{0pt}
\setlength{\belowdisplayskip}{0pt}
\setlength{\itemsep}{0pt}
\setlength{\partopsep}{0pt}
}

\pagenumbering{gobble}

\title{Auto-Validate: Unsupervised Data Validation \\ Using Data-Domain Patterns Inferred from Data Lakes}

\author{Jie Song}
\authornote{Work done at Microsoft Research.}
\affiliation{\institution{University of Michigan}}
\email{jiesongk@umich.edu}

\author{Yeye He}
\affiliation{\institution{Microsoft Research}}
\email{yeyehe@microsoft.com}

\begin{abstract}
Complex data pipelines are increasingly common
in diverse applications such as BI reporting and
ML modeling. These pipelines 
often recur regularly (e.g., daily or weekly), 
as BI reports need to be refreshed, 
and ML models need to be retrained. However, it is
widely reported that in complex production
pipelines, upstream data feeds can change in 
unexpected ways, causing downstream applications 
to break silently that are expensive to resolve.

Data validation has thus become an important topic,
as evidenced by notable recent efforts from Google and Amazon,
where the objective is to catch data quality 
issues early as they arise in the pipelines.
Our experience on production data suggests, however, 
that on string-valued data, these existing approaches yield high
false-positive rates and frequently require human intervention.
In this work, we develop a corpus-driven approach to
auto-validate \emph{machine-generated data} by
inferring suitable data-validation ``patterns''
that accurately describe the underlying data-domain,
which minimizes false-positives while
maximizing data quality issues caught.
Evaluations using production
data from real data lakes
suggest that \sj{} is substantially more effective
than existing methods. Part of this 
technology ships as 
an \textsc{Auto-Tag} feature in \textsc{Microsoft Azure Purview}.
\end{abstract}

\maketitle

\vspace{-2mm}
\section{Introduction}

Complex data pipelines and data flows are increasingly common
in today's BI, ETL and ML systems
(e.g., in Tableau~\cite{tableau-flow}, 
Power BI~\cite{power-bi-flow}, Amazon Glue~\cite{amazon-workflow},
Informatica~\cite{Informatica}, Azure ML~\cite{azureml-pipelines}, etc.).
These pipelines typically \textit{recur regularly} (e.g., daily), 
as BI reports need to be refreshed regularly~\cite{dayal2009data, schelter2018automating}, 
data warehouses need to be 
updated frequently~\cite{vassiliadis2009near}, and
ML models need to be retrained continuously (to
prevent model degradation)~\cite{breck2019data, polyzotis2017data}.

However, it is
widely recognized (e.g.,~\cite{breck2019data, hynes2017data, polyzotis2017data, schelter2018automating, 
schelter2019unit, schelter2019differential}) 
that in recurring production
data pipelines, over time upstream data feeds can often change in
unexpected ways. For instance, over time, data
columns can be added to or removed in upstream data,
creating \textit{schema-drift}~\cite{breck2019data, 
polyzotis2017data,  schelter2018automating}. 
Similarly, \textit{data-drift}~\cite{polyzotis2017data} is also
common, as the 
formatting standard of data values can change silently
(e.g., from ``\val{en-us}'' to ``\val{en-US}'', as reported in~\cite{polyzotis2017data}); and
invalid values
can also creep in (e.g., ``\val{en-99}'', for unknown locale).

Such schema-drift and data-drift can lead to quality issues
in downstream applications, which are both hard to detect 
(e.g., modest model degradation due to  
unseen data~\cite{polyzotis2017data}), and
hard to debug (finding root-causes in complex pipelines
can be difficult).
These silent failures require substantial human
efforts to resolve, 
increasing the cost of running production data pipelines
and decreasing the effectiveness of data-driven 
system~\cite{breck2019data, polyzotis2017data, 
schelter2018automating, schelter2019unit}.

\textbf{Data Validation by Declarative Constraints.}
Large tech companies operating large numbers of complex pipelines are the first to recognize the need to catch data quality issues
early in the pipelines. In response, they pioneered a number of 
``data validation'' tools, including Google's TensorFlow Data 
Validation (TFDV)~\cite{breck2019data, TFDV} and
Amazon's Deequ~\cite{Deequ, schelter2018automating}, etc.
These tools develop easy-to-use \textit{domain specific languages} (DSLs),
for developers and data engineers to write declarative
\textit{data constraints} that describe how ``normal'' data 
should look like in the pipelines, so that
any unexpected deviation introduced over time can be caught early
as they arise.

\begin{figure}[t]
        \centering
        \includegraphics[width=0.45\textwidth]{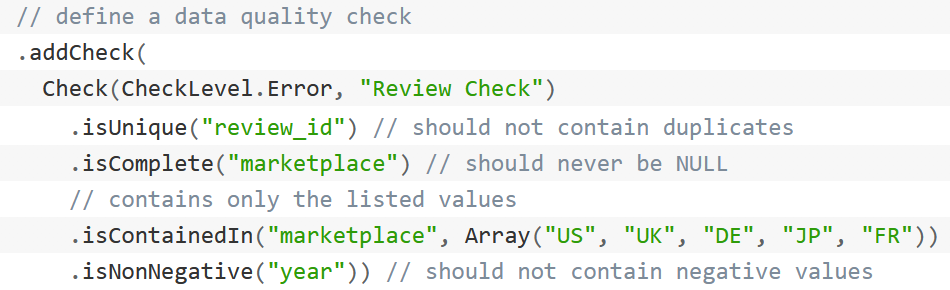}
\caption{Example code snippet to describe expected data-values, using declarative constraints in Amazon Deequ.}
\vspace{3mm}
\label{fig:deequ}
\end{figure}

Figure~\ref{fig:deequ} shows an example code snippet\footnote{\url{https://aws.amazon.com/blogs/big-data/test-data-quality-at-scale-with-deequ/}}
from Amazon Deequ~\cite{schelter2018automating} to specify
data validation constraints on a table. 
In this case, it is declared
that values in the column ``\val{review\_id}'' need to be unique, column
``\val{marketplace}'' is complete (with no
\val{NULL}s), values in ``\val{marketplace}'' need to be
in a fixed dictionary 
\{``\val{US}'', ``\val{UK}'', ``\val{DE}'', ``\val{JP}'', ``\val{FR}''\}, etc.
Such constraints will be used to validate against data arriving
in the future, and raise alerts if violations are detected.
Google's TFDV uses a similar declarative approach 
with a different syntax.

These approaches allow users to
provide \textit{high-level specifications} of data
constraints, and is a significant improvement over 
 \textit{low-level} assertions used in ad-hoc validation scripts (which are
are hard to program and maintain)~\cite{breck2019data, TFDV,
Deequ, schelter2018automating, swami2020data}.


\textbf{Automate Data Validation using Inferred Constraints.}
While declarative data validation clearly improves over 
ad-hoc scripts, and is beneficial as reported
by both Google and Amazon~\cite{breck2019data, schelter2018automating},
the need for data engineers to manually write
data constraints \textit{one-column-at-a-time} (e.g., 
in Figure~\ref{fig:deequ}) still makes it time-consuming 
and hard to scale. This is especially true for complex production data,
where a single data feed can have hundreds of
columns requiring substantial manual work.
As a result, experts today can only afford to write 
validation rules for
part of their data (e.g., important columns). 
Automating the generation of 
data validation constraints has 
become an important problem.

The main technical challenge here is that validation rules 
have to be inferred using data values 
\textit{currently observed} from a 
pipeline, but the inferred rules have to apply
to future data that \textit{cannot yet be observed}
(analogous to training-data/testing-data in ML settings where
test-data cannot be observed).
Ensuring that validation rules \textit{generalize} to unseen future data is thus
critical.

\begin{figure*}
        \centering
        \includegraphics[width=0.9\textwidth]{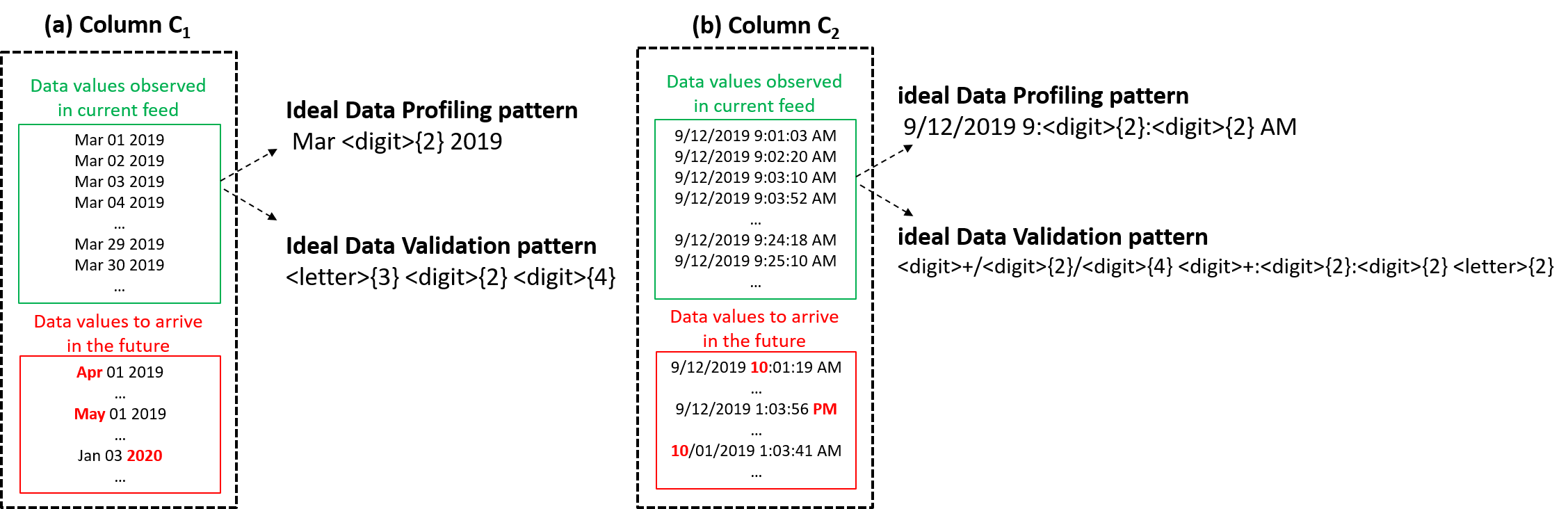}
\caption{Two example data columns $C_1$ and $C_2$. 
The top portion of each column
shows data in the current feed that can be observed, 
while the bottom portion shows
values that will arrive in the future.}
\label{fig:datetime-example}
\end{figure*}

We should note that existing solutions already 
make efforts in this direction -- 
both Google TFDV and Amazon Deequ can scan 
an existing data feed, and automatically suggest validation
constraints for human engineers to inspect and approve.
Our analysis suggests that the capabilities of
these existing solutions are still rudimentary, especially
on \textit{string-valued data columns}.
For instance, for the example $C_1$
in Figure~\ref{fig:datetime-example}(a) with
date strings in \val{Mar 2019},
TFDV would infer a validation rule\footnote{This was initially tested on TFDV version 0.15.0, the latest version available in early 2020. We tested it again at the time of publication in March 2021 using the latest TFDV version 0.28.0, and observed the same
behavior for this test case.}  requiring all future values
in this column to come from a fixed dictionary 
with existing values in $C_1$:
\{`` \val{Mar 01 2019}'', \ldots `` \val{Mar 30 2019}'' \}.
This is too restrictive for data validation, 
as it can trigger false-alarms on future data
(e.g. values like ``\val{Apr 01 2019}'').
Accordingly, TFDV documentation does not recommend
suggested rules to be used directly, and
``\textit{strongly advised developers to review 
the inferred rule and refine it as needed}''\footnote{\url{https://www.tensorflow.org/tfx/data_validation/get_started}}.

When evaluated on production data (without human intervention), we
find TFDV produces false-alarms on over 90\%
string-valued columns.
Amazon's Deequ considers distributions of values
but still falls short, triggering false-alarms on over 20\% columns.
Given that millions of data columns are processed daily
in production systems, these false-positive rates
can translate to millions of false-alarms,
which is not adequate for automated applications.

\textbf{Auto-Validate string-valued data using patterns.}
We in this work aim to infer
accurate data-validation rules for \textit{string-valued data},
using regex-like patterns. We perform an in-depth
 analysis of production data 
in real pipelines, from a large 
Map-Reduce-like system used 
at Microsoft~\cite{patel2019big, zhou2012scope},
which hosts over 100K production jobs each 
day, and powers popular products like Bing Search and Office.
We crawl a sample of 7M columns from the data lake hosting
these pipelines and find string-valued columns
to be prevalent, accounting for 75\% of 
columns sampled.

\begin{figure*}[t]
        \centering
        \includegraphics[width=1\textwidth]{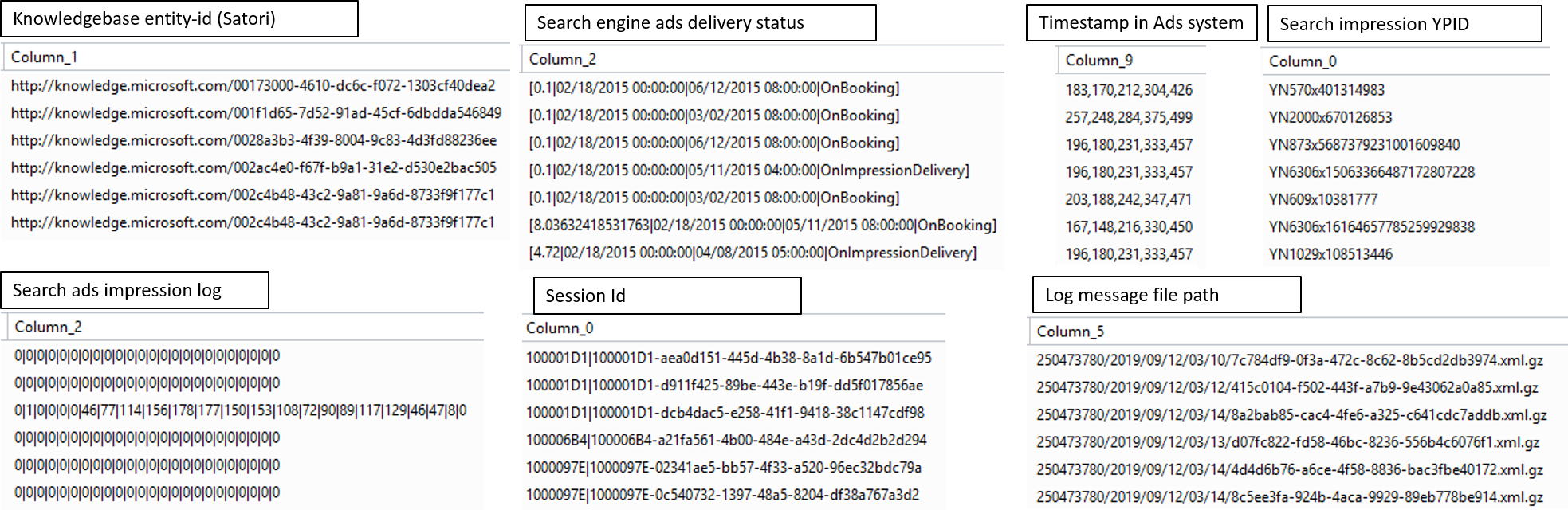}
\caption{Example columns crawled from a production enterprise data lake.  Each
column has a distinctive data pattern in proprietary
formats, encoding specific meanings (the first column is
anonymized for review). 
These are all common patterns found in at least 5000 columns.}
\label{fig:domain-example}
\vspace{-3mm}
\end{figure*}

Figure~\ref{fig:domain-example} shows 7 example string-valued
columns from the crawl. A key characteristic common to
many such columns, is that these columns have \textit{homogeneous}
and \textit{machine-generated} data values, which exhibit
distinct ``patterns'' that encode specific semantic meanings. In our 
examples,  these include 
knowledge-base entity-id~\cite{BingEntity} (used by Bing Search), 
online-ads delivery status (used by search ads), 
time-stamps in proprietary formats  (used by search ads), etc.
We note that these are not one-off patterns occurring just
once or twice -- each pattern 
in Figure~\ref{fig:domain-example}
shows up in at least 5000 data columns in our crawl, suggesting
that these are widely-used concepts.
Similar data are likely common in other industries and
domains (e.g., for pharmaceutical or
financial companies~\cite{stonebraker2018data}).

In this work, we focus on these string-valued columns
that have homogeneous machine-generated data values.
Specifically, if we could infer suitable patterns to describe the
underlying ``domain'' of the data (defined as the space of 
all valid values),  we can use such patterns 
as validation rules against data in the future.
For instance, the pattern 
``{\small \texttt{<letter>\{3\} <digit>\{2\} <digit>\{4\}}}''
can be used to validate 
$C_1$ in Figure~\ref{fig:datetime-example}(a),
and ``{\small \texttt{<digit>+/<digit>\{2\}/<digit>\{4\} <digit>+:<digit>\{2\}:<digit>\{2\} <letter>\{2\}}}''
for $C_2$ in Figure~\ref{fig:datetime-example}(b).

Our analysis on a random sample of 1000 columns 
from our production data lake
suggests that around 67\% string-valued columns
have homogeneous machine-generated values
(like shown in Figure~\ref{fig:domain-example}), whose
underlying data-domains are amenable to pattern-based
representations (the focus of this work). 
The remaining 33\% columns have \textit{natural-language (NL) content}
(e.g., company names, department names, etc.),
for which pattern-based approaches would be less well-suited.
We should emphasize that for large data-lakes and production
data pipelines, machine-generated data will likely dominate (because 
machines are after all more productive at churning out large volumes of data
than humans producing NL data), making our proposed
technique widely applicable.

\iftoggle{fullversion}{
While on the surface this looks similar 
to prior work on \textit{pattern-profiling}
(e.g., Potter's Wheel~\cite{raman2001potter}
and PADS~\cite{fisher2005pads}), which
also learn patterns from values, 
we highlight the key differences in their objectives
that make the two entirely different problems.

\textbf{Data Profiling vs. Data Validation.}
In pattern-based data profiling, the main objective is to ``summarize''
a large number of values in a column $C$, 
so that users can quickly understand
content is in the column without needing to scroll/inspect every value.
As such, the key criterion is to select 
patterns that can succinctly 
describe given data values in $C$ \textit{only},
\textit{without needing to consider values not present in $C$}.

For example, existing pattern profiling
methods like Potter's Wheel~\cite{raman2001potter}
 would correctly
generate a desirable pattern 
``{\small \texttt{Mar <digit>\{2\} 2019}}'' for $C_1$ in 
Figure~\ref{fig:datetime-example}(a), which is valuable
from a pattern-profiling's perspective as it succinctly describes
values in  $C_1$.
However, this pattern is not suitable for 
data-validation, as it is overly-restrictive
and would trigger false-alarms for values
like ``{\small \texttt{Apr 01 2019}}'' arriving in the future. A more appropriate
data-validation pattern should instead be
``{\small \texttt{<letter>\{3\} <digit>\{2\} <digit>\{4\}}}''. 
Similarly, for $C_2$, Figure~\ref{fig:datetime-example}(b) shows
that a pattern ideal for profiling
is again very different from one suitable for data-validation.
}

\begin{figure}
        \centering
        \includegraphics[width=0.3\textwidth]{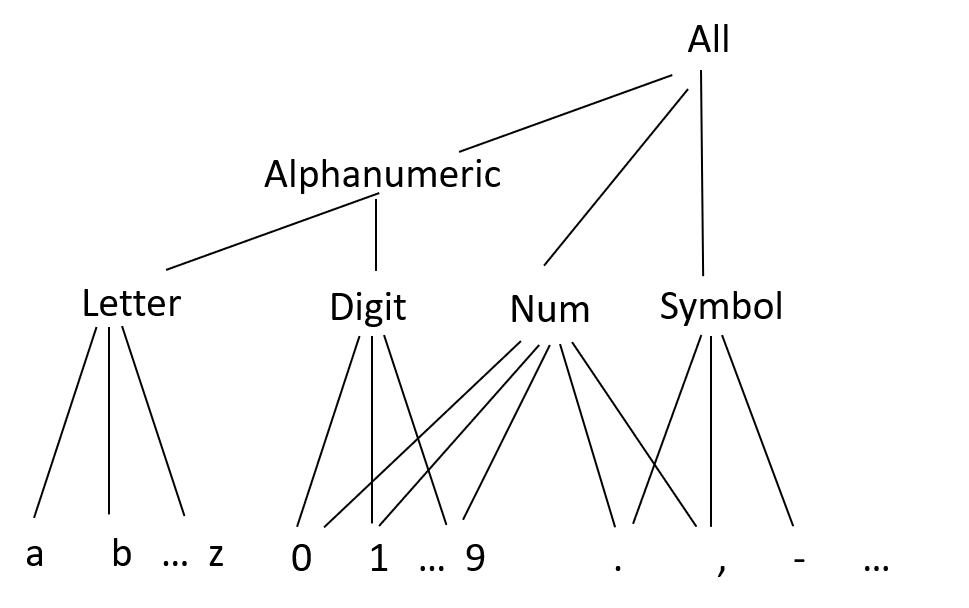}
\caption{Example string generalization hierarchy.}
\label{fig:hierarchy}
\end{figure}

\begin{figure}
        \centering
        \includegraphics[width=0.5\textwidth]{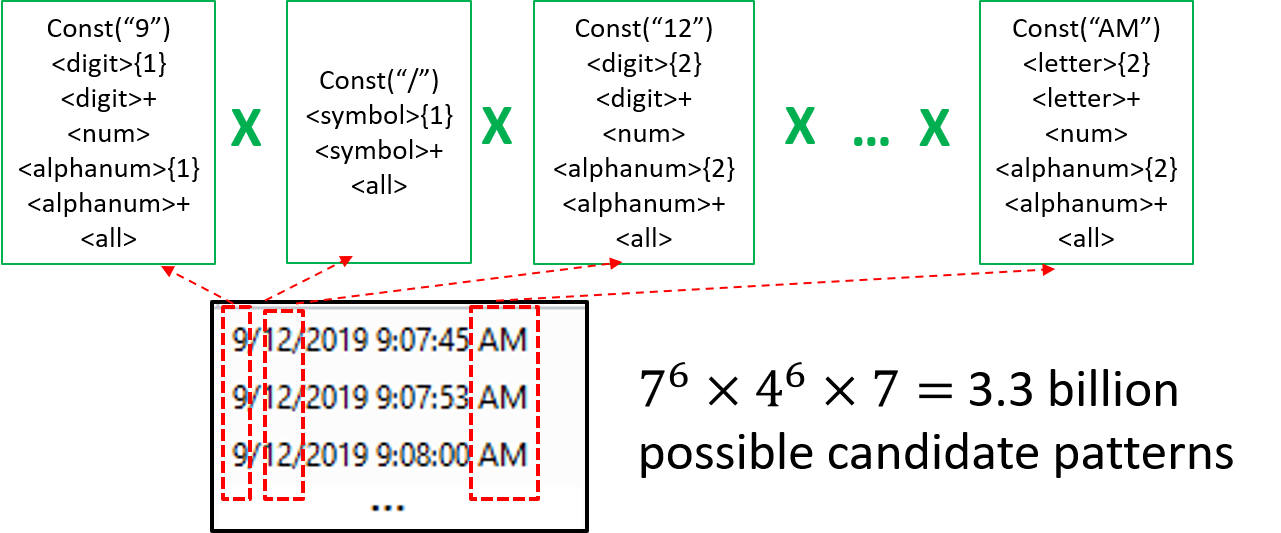}
\caption{Possible ways to generalize a column of date-time strings, using the
hierarchy in Figure~\ref{fig:hierarchy}.}
\label{fig:generalization}
\end{figure}

Conceptually, there are many ways to ``generalize'' a column
of data into patterns. For a simple column of date-time strings
like in Figure~\ref{fig:generalization}, and using a standard generalization 
hierarchy (Figure~\ref{fig:hierarchy}), one could generate over 3 billion
patterns. For example, the first part
(digit ``\val{9}'' for month) alone can be generalized in 7 different
ways below, also shown in the top-right box of Figure~\ref{fig:generalization}.
\begin{small}
\begin{itemize}[leftmargin=*, ,topsep=0pt]
\item[](1) a constant ``{\small \texttt{9}}'' ({\small \texttt{Const(``9'')}}), 
\item[](2) one single digit ({\small \texttt{<digit>\{1\}}}), 
\item[](3) at least one digits ({\small \texttt{<digit>+}}), 
\item[](4) any number including floating-points ({\small \texttt{<num>}}), 
\item[](5) one alpha-numeric character ({\small \texttt{<alphanum>}}), 
\item[](6) at least one alpha-numeric characters ({\small \texttt{<alphanum>+}}), 
\item[](7) root of the hierarchy ({\small \texttt{<all>}})
\end{itemize}
\end{small}

The cross-product of these options at each position 
creates a large space of patterns (about 3.3 billion)  
  for this simple column.

The challenge for data-validation, is to select suitable
patterns from this large space given only
observed values in $C$, in
anticipation of valid values from the same ``domain''
that may arrive in the future.
The key is to not use
overly restrictive patterns (as in data-profiling),
which would yield many false-positive alerts.
Conversely overly general 
patterns (like the trivial ``\texttt{.*}'') should also
not be used, as they would 
fail to detect any data quality issues. 
In effect, we want to find
patterns that balance two conflicting goals: 
(1) reduce false-alarms 
(or improve detection precision); 
and (2) catch as many issues
as possible (or improve detection recall).

\textbf{A corpus-driven approach to pattern-inference.}
Intuitively, the pattern-inference problem is hard if we 
only look at one input column $C$ -- for the date-time examples in 
Figure~\ref{fig:datetime-example}, 
we as humans know what
patterns are suitable; 
but for the examples in Figure~\ref{fig:domain-example}
drawn from proprietary domains, even
humans may find it hard to determine suitable validation
patterns, and would need to seek
``additional evidence'', such as
other ``similar-looking columns''.  

Following this intuition, we in this work formulate
pattern-inference as optimization problems, by
leveraging a large corpus of related tables 
$\mathbf{T}$ (e.g. data produced by production pipelines and
dumped in
the same enterprise data lake). Our inference algorithms leverage 
columns ``similar'' to $C$ in $\mathbf{T}$, to 
reason about patterns that may be ideal for data-validation.

Large-scale evaluations on 
production data suggest that \sj{} produces 
constraints that are substantially more accurate than existing 
methods. Part of this technology ships as an \textsc{Auto-Tag} feature 
in \textsc{Microsoft Azure Purview}~\cite{purview}.

\vspace{-2mm}
\section{Auto-Validate (Basic Version)}

We start by describing a basic-version of \sj{} for ease of illustration,
before going into more involved
algorithm variants that use vertical-cuts (Section~\ref{sec:vertical-cuts})
and horizontal-cuts  (Section~\ref{sec:horizontal-cuts}).

\vspace{-2mm}
\subsection{Preliminary: Pattern Language}
\label{sec:language}
Since we propose to validate data by patterns,
we first briefly describe the pattern language used. 
We note that this is fairly standard (e.g., similar to 
ones used in~\cite{raman2001potter}) 
and not the focus of this work.

Figure~\ref{fig:hierarchy} shows a generalization hierarchy 
used, where leaf-nodes represent the English alphabet, 
and intermediate nodes (e.g., 
\val{<digit>}, \val{<letter>}) represent \textit{tokens} that values
can generalize into. A pattern is simply a sequence 
of (leaf or intermediate) tokens,
and for a given value $v$, this hierarchy
induces a space of all patterns consistent with $v$, 
denoted by $\mathbf{P}(v)$.
For instance, for a value $v =$ ``\val{9:07}'', we can generate
$\mathbf{P}(v) = $ \{
 ``\val{<digit>:<digit>\{2\}}'', 
 ``\val{<digit>+:<digit>\{2\}}'', 
 ``\val{<digit>:<digit>+}'', 
 ``\val{<num>:<digit>+}'',  ``\val{9:<digit>\{2\}}'', \ldots
\}, among many other options.

Given a column $C$ for which patterns need
to be generated, we define the 
space of \textit{hypothesis-patterns}, 
denoted by $\mathbf{H}(C)$, as the set of patterns 
consistent with all values  $v \in C$, or 
$\mathbf{H}(C) = \cap_{v \in C}{\mathbf{P}(v)} \setminus \text{``.*''}$ (note that we exclude the trivial ``.*'' pattern as it 
is trivially consistent with all $C$).
Here we make an implicit assumption that values in 
$C$ are homogeneous and drawn from the same domain, which is generally
true for \textit{machine-generated data} from production
pipelines (like shown in 
Figure~\ref{fig:domain-example})\footnote{Our empirical sample
in an enterprise data lake suggests that 87.9\% columns are 
homogeneous (defined as drawn from the same underlying
domain), and 67.6\% are homogeneous with
machine-generated patterns.}.
Note that the assumption is used to only simplify our discussion of 
the basic-version of \sj (this section), and will later be relaxed
in more advanced variants (Section~\ref{sec:horizontal-cuts}).

 We call $\mathbf{H}(C)$ hypothesis-patterns because as we will see, 
each $h(C) \in \mathbf{H}(C)$
will be tested like a ``hypothesis'' to determine if $h$ is a good
validation pattern.  
We use $h$ and $h(C)$
interchangeably when the context is clear. 
\begin{example}
\label{ex:patterns}
Given the column $C$ shown in Figure~\ref{fig:generalization},
and the hierarchy in Figure~\ref{fig:hierarchy}, the 
space of hypothesis patterns $\mathbf{H}(C)$ is shown 
at the top of  Figure~\ref{fig:generalization}.
\end{example}

We use an in-house implementation to produce patterns based
on the hierarchy, which can be considered
as a variant of well-known pattern-profiling
techniques like~\cite{raman2001potter}. Briefly,
our pattern-generation works in two steps. In the
first step, it scans each cell value and
emits \textit{coarse-grained} tokens 
(e.g., \val{<num>} and \val{<letter>+}), 
without specializing into 
\textit{fine-grained} tokens 
(e.g., \val{<digit>\{1\}} or \val{<letter>\{2\}}).
For each value in Figure~\ref{fig:generalization},
for instance, this produces the same 
\val{<num>/<num>/<num> <num>:<num>:<num> <letter>+}.
Each coarse-grained pattern is then checked for coverage
so that only patterns meeting a coverage threshold are retained. 
In the second step, each coarse-grained pattern is
specialized into fine-grained patterns (examples are shown in 
Figure~\ref{fig:generalization}), as long as the given
threshold is met. 
\iftoggle{fullversion}
{Pseudo-code of the procedure can be found in 
Algorithm~\ref{algo:pattern} below.}
{More details of this can be found in a 
full version of this paper~\cite{full}. 
}

\iftoggle{fullversion}
{
\begin{algorithm}
\SetAlgoLined
  \SetKwInOut{Input}{input}
  \SetKwInOut{Output}{output}
  \SetKwData{S}{$\mathcal{S}$}
  \SetKwData{P}{$\mathcal{P}$}
  \SetKwProg{GeneratePatterns}{GeneratePatterns}{}{}

  \GeneratePatterns{$(\S, H)$}{
  \Input{string values $\S = \{s_1, \dots, s_n\}$, generalization hierarchy $H$}
  \Output{patterns $\P_{fine}$ of $\S$ induced by $H$}
$\P_{coarse} \leftarrow \{\}$\;
\ForEach{$s \in \S$ }{
p $\leftarrow$ GenerateCoarsePatterns(s)\;
add p to $\P_{coarse}$
}
retain patterns in $\P_{coarse}$ with sufficient coverage\;
$\P_{fine} \leftarrow \{\}$\;
\ForEach{$p$ $\in$ $\P_{coarse}$}{
p' $\leftarrow$ drill-down each $p$ using $H$\;
add p' to $\P_{fine}$ if coverage is sufficient \;
}
   \KwRet{$\P_{fine}$}\;
}
\caption{Generate patterns from a column of values}\label{algo:pattern}
\end{algorithm}
}
{}

We emphasize that the
proposed \sj{} framework is not tied to specific 
choices of hierarchy/pattern-languages -- in fact, it is 
entirely orthogonal to such choices, as other variants of languages 
are equally applicable in this framework.

\vspace{-2mm}
\subsection{Intuition: Goodness of Patterns}
\label{sec:intuition}
The core challenge in  \sj{}, is to determine whether a pattern
$h(C) \in \mathbf{H}(C)$ is  suitable for data-validation.
Our idea here is to leverage a corpus of 
related tables $\mathbf{T}$ (e.g., drawn from the
same enterprise data lake, which stores input/output data
of all production pipelines). Intuitively, a pattern
$h(C) \in \mathbf{H}(C)$ is a ``good'' validation
pattern for $C$, if it accurately describes the underlying ``domain'' of $C$,
defined as the space of all valid data values.
Conversely, $h(C)$ is a ``bad'' validation pattern, if:
\begin{itemize}[leftmargin=*,topsep=0pt]
\item[](1) $h(C)$ does not capture all valid values in the underlying ``domain'' of $C$;
\item[](2) $h(C)$ does not produce sufficient matches in $\mathbf{T}$.
\end{itemize}

Both are intuitive requirements -- 
(1) is desirable, because as we discussed, the danger in data-validation is
to produce patterns 
that do not cover all valid values from the same domain, 
which leads to false alarms.
Requirement (2) is also sensible because we want to see sufficient
evidence in $\mathbf{T}$ 
before we can reliably 
declare a pattern $h(C)$ to be a common ``domain''
of interest (seeing a pattern once or twice is not sufficient).

We note that both of these two requirements can be reliably tested using 
$\mathbf{T}$ alone and without involving humans.
Testing (2) is straightforward using standard pattern-matches. 
Testing (1) is also feasible,
under the assumption that most data columns 
in $D \in \mathbf{T}$ contain homogeneous
values drawn from the same underlying 
domain, which is generally true for ``machine-generated'' data
from production pipelines.
(Homogeneity is less likely to hold on other types of data, e.g., data manually typed in by human operators).

We highlight the fact that
we are now dealing with two types of columns that
can be a bit confusing --
the first is the input column $C$ for which we need to
generate validation patterns $h$, and the second is 
columns $D \in \mathbf{T}$ from which we draw evidence 
to determine the goodness of $h$.
To avoid confusion,  we will follow the standard of referring
to the first as the \textit{query column}, denoted by $C$,
and the second as \textit{data columns}, denoted by $D$.

\begin{figure*}
        \centering
        \includegraphics[width=0.95\textwidth]{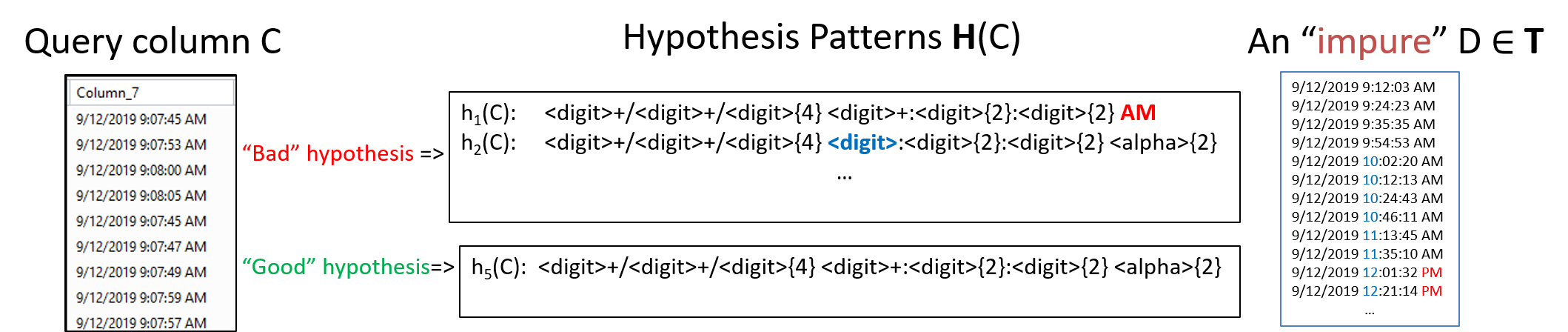}
\caption{Leverage corpus $\mathbf{T}$ to determine ``good'' vs. ``bad'' hypothesis patterns, for a given query column $C$.}
\vspace{-3mm}
\label{fig:intuition}
\end{figure*}

We now show how to test requirement (1) using $\mathbf{T}$ below.

\begin{example}
\label{ex:intuition} 
The left of
Figure~\ref{fig:intuition} shows a query column $C$ for which
validation patterns need to be generated. A few
hypothesis-patterns in $\mathbf{H}(C)$ are listed in
the middle. In this example, we intuitively know
that $h_1(C)$ and $h_2(C)$ are ``bad'' patterns
since they 
are too ``narrow'' and would lead to false-alarms,
while $h_5(C)$ is a ``good'' validation pattern
that suitably generalizes the underlying domain. 

We show that such determinations can 
be inferred using $\mathbf{T}$.
Specifically, $h_1(C)$ is a ``bad'' pattern based on $\mathbf{T}$,
because we can find many columns like $D \in \mathbf{T}$ 
shown on the right of Figure~\ref{fig:intuition}
that are ``impure'' -- these columns contain
values that match $h_1(C)$, as well as values that do not
(e.g., ``{\small \texttt{9/12/2019 12:01:32 PM}}'',
where the ``{\small \texttt{PM}}'' part does not match $h_1(C)$).
A large number of ``impure'' columns like this would
indicate that $h_1(C)$ is not a good pattern to describe 
the underlying domain because if it were, we should not see
many impure columns like $D \in \mathbf{T}$ given homogeneity.

Similarly, we can show that $h_2(C)$ is not a good validation pattern,
for using $h_2(C)$ to describe the 
domain would again make many columns like $D$ ``impure'' 
(values like ``\val{10:02:20 AM}'' are inconsistent
with $h_2(C)$ since they have two-digit hours, whereas
$h_2(C)$ uses a single \val{<digit>} for hours),
  suggesting $h_2(C)$ is also too narrow.

Using the ideal pattern $h_5(C)$ to describe the domain, 
on the other hand, would yield few
``impure'' columns in $\mathbf{T}$.
\end{example}

Intuitively, we can use the \textit{impurity} of pattern $h$ on
data columns $D \in \mathbf{T}$,
to infer whether $h$ is a good validation pattern
for the domain, defined as follows.
\begin{definition}
\label{def:impurity}
The \textit{impurity} of a data column 
$D \in \mathbf{T}$, relative to a given hypothesis pattern $h$,
is defined as:
\begin{equation}
\label{eqn:impurity}
\text{Imp}_{D}(h) = \frac{|\{v | v \in D, h \notin \mathbf{P}(v)\}|}{|\{ v | v \in D \}|}
\end{equation}
\end{definition}

This definition is intuitive -- we measure 
impurity as the fraction 
of values in $D$ not matching $h$.

\begin{example}
\label{ex:fpr}
In Figure~\ref{fig:intuition}, 
$\text{Imp}_D(h_1)$ can be calculated as $\frac{2}{12}$,
since the last 2 values (with ``\val{PM}'') 
out of 12 do not match $h_1$.

Similarly, $\text{Imp}_D(h_2)$ can be calculated as $\frac{8}{12}$,
since the last 8 values in $D$ (with two-digit hours) 
do not match $h_2$.

Finally, $\text{Imp}_D(h_5)$ is $\frac{0}{12}$,
since all values in $D$ match $h_5$.
\end{example}

We note that if $h(C)$ is used to validate
a data column $D$ (in the same domain as $C$)
arriving in the future, then 
Imp$_D(h)$ directly corresponds to
expected false-positive-rate (FPR):

\begin{definition}
\label{def:fpr}
The expected \textit{false-positive-rate} (FPR) of using pattern $h(C)$ 
to validate a data column $D$ drawn from the same domain as $C$,
denoted by $\text{FPR}_{D}(h)$, is defined as:
\begin{equation}
\label{f}
\text{FPR}_{D}(h) = \frac{\text{FP}_{D}(h)}{\text{TN}_{D}(h)+\text{FP}_{D}(h)}
\end{equation}
Where FP$_{D}(h)$ and TN$_{D}(h)$ are the numbers of 
false-positive detection and true-negative detection of $h$ on $D$,
respectively. 
\end{definition}
Note that since $D$ is from the same domain as $C$,
ensuring that TN$_{D}(h)$ + FP$_{D}(h)$ = $|D|$, 
$\text{FPR}_{D}(h)$ can be rewritten as:
\begin{equation}
\label{eqn:fpr}
\text{FPR}_{D}(h) = \frac{|\{v | v \in D, h \notin \mathbf{P}(v)\}|}{|\{v | v \in D \}|} = \text{Imp}_D(h)
\end{equation}
Thus allowing us to estimate $\text{FPR}_{D}(h)$ using $\text{Imp}_D(h)$.

\begin{example}
\label{ex:fpr-impurity}
Continue with Example~\ref{ex:fpr}, it can be
verified that the expected 
FPR of using $h$ as the validation pattern for $D$,
directly corresponds to the impurity $\text{Imp}_D(h)$ --
e.g., using $h_1$ to validate $D$ has $\text{FPR}_{D}(h_1)$
= $\text{Imp}_D(h_1)$ = $\frac{2}{12}$; while using 
$h_5$ to validate $D$ has $\text{FPR}_{D}(h_5) = \text{Imp}_D(h_5) = 0$, etc.
\end{example}

Based on FPR$_{D}(h)$ defined for one column $D \in \mathbf{T}$, 
we can in turn define the estimated FPR on the entire corpus $\mathbf{T}$,
using all column $D \in \mathbf{T}$ 
where some value $v \in D$ matches $h$:
\begin{definition}
\label{def:fpr_corpus}
Given a corpus $\mathbf{T}$, we estimate the 
FPR of pattern $h$ on  $\mathbf{T}$, 
denoted by FPR$_{\mathbf{T}}(h)$, as:
\begin{equation}
\label{eqn:fpr_agg}
\text{FPR}_{\mathbf{T}}(h) = \avg_{D \in \mathbf{T}, v \in D, h \in \mathbf{P}(v) }{\text{FPR}_{D}(h)}
\end{equation} 
\end{definition}
\vspace{-2mm}


\begin{example}
\label{ex:fpr_t}
Continue with the $h_5$ in Example~\ref{ex:fpr-impurity}. Suppose 
there are 5000 data columns $D \in \mathbf{T}$ 
where some value $v \in D$ matches $h_5$.
Suppose 4800 columns out of the 5000 have $\text{FPR}_{D}(h_5) = 0$, 
and the remaining 200 columns have $\text{FPR}_{D}(h_5) = 1\%$. 
The overall $\text{FPR}_{\mathbf{T}}(h_5)$ 
can be calculated as $\frac{200 * 1\%}{5000} = 0.04\%$,
using Equation~\eqref{eqn:fpr_agg}.
\end{example}

We emphasize that a low FPR is 
critical for data-validation, because given a large 
number of columns in production 
systems, even a 1\% FPR
translates into many false-alarms that require human
attention to resolve. FPR is thus a key objective we minimize.

\vspace{-2mm}
\subsection{Problem Statement}
We now describe the basic version 
of \sj{} as follows. 
Given an input query column $C$ and a background corpus $\mathbf{T}$,
we need to produce a validation pattern $h(C)$, 
such that $h(C)$ is expected to have a low FPR
but can still catch many quality issues.
We formulate this as an optimization problem.
Specifically, we introduce an FPR-Minimizing 
version of Data-Validation (FMDV), defined as:
\begin{align}
\hspace{-1cm} \text{(FMDV)} \qquad{} \min_{h \in \mathbf{H}(C)} & \text{FPR}_\mathbf{T}(h)   \label{eqn:fmdv_obj} \\ 
 \mbox{s.t.} ~~ & \text{FPR}_\mathbf{T}(h) \leq r \label{eqn:fmdv_fpr} \\
 & \text{Cov}_\mathbf{T}(h) \geq m \label{eqn:fmdv_cov}
\end{align}
Where $\text{FPR}_\mathbf{T}(h)$ is the 
expected FPR of $h$ estimated using $\mathbf{T}$,
which has to be lower than a given target threshold $r$
(Equation~\eqref{eqn:fmdv_fpr}); and
$\text{Cov}_\mathbf{T}(h)$ is the coverage of $h$, or 
the number of columns in $\mathbf{T}$ that match $h$,
which has to be greater than a given threshold 
$m$ (Equation~\eqref{eqn:fmdv_cov}). Note that
these two constraints directly map to the 
two ``goodness'' criteria in Section~\ref{sec:intuition}.

The validation pattern $h$ we produce for $C$ is
then the minimizer
of FMDV from the hypothesis space 
$\mathbf{H}(C)$ (Equation~\eqref{eqn:fmdv_obj}).
We note that this can be interpreted as a conservative approach
that aims to find a ``safe'' pattern $h(C)$ with minimum FPR.


\begin{example}
\label{ex:fmdv}
Suppose we have FMDV with a target FPR rate $r$ no greater
than $0.1\%$, and a required coverage of at least $m = 100$.
Given the 3 example hypothesis patterns in Figure~\ref{fig:intuition},
we can see that patterns $h_1$ and $h_2$ are not feasible solutions
because of their high estimated FPR.
Pattern $h_5$ has an estimated FPR of 0.04\% and a coverage of 5000,
which is a feasible solution to FMDV.
\end{example}

\begin{lemma}
\label{lem:fmdv}
In a simplified scenario where each 
column in $\mathbf{T}$ has values drawn randomly from
exactly one underlying domain, and each domain in turn 
corresponds to one ground-truth
pattern. Then given a query column $C$ for which a validation pattern
needs to be generated,
and a corpus $\mathbf{T}$ with a large number ($n$) of 
columns generated from the same domain as $C$, 
with high probability (at least $1 - (\frac{1}{2})^n$), 
the optimal solution of FMDV 
for $C$ (with $r=0$) converges to the ground-truth pattern of $C$.
\end{lemma}

\iftoggle{fullversion}{
\begin{proofsketch}
We show that any candidate patterns 
that ``under-generalizes'' the
domain will with high probability 
be pruned away due to FPR violations (given $r=0$).
To see why this is the case, consider a pattern $p$ produced for $C$
that under-generalizes the ground-truth pattern $g$. Note
$p$ is picked over $g$ only when not a single value in all $n$ columns 
in the same domain as $C$
drawn from the space of values defined 
by $g$ are not in $p$ (since $r$ is set to 0). 
Because these $n$ columns are drawn randomly
from $g$ and each column has at least one value,
the probability of this happening is thus at most $(\frac{1}{2})^n$
($\frac{1}{2}$ is used because there is at least one branch
$p'$ in $g$ that is not in $p$), making the overall probability of success to be
$1 - (\frac{1}{2})^n$.
\end{proofsketch}
}
{
Intuitively, this result holds under the simplifying assumptions, 
because when given a sufficiently large number of $n$ columns in
$\mathbf{T}$ from the same domain as $C$, any candidate patterns 
that ``under-generalizes'' the
domain will be pruned away due to FPR violations, 
leaving the ground-truth pattern to be the optimal solution.
More details of the result can be found in~\cite{full}.
}

We also explored alternative formulations, such as 
coverage-minimizing data-validation (CMDV), where
we minimize coverage $\text{Cov}_\mathbf{T}(h)$
in the objective function in place of FPR$_\mathbf{T}(h)$ as in FMDV.
We omit details in the interest of space, 
but will report that the conservative FMDV is more
effective in practice. 

A dual version of FMDV can be used for automated data-tagging, 
where we want to find the most restrictive 
(smallest coverage) pattern
to describe the underlying domain (which can be used to ``tag'' related
columns of the same type), under the constraint
of some target false-negative rate. We describe this dual version
in~\cite{auto-tag-tr}.

\subsection{System Architecture}
\label{sec:architecture}
While FMDV is simple overall, 
a naive implementation would
require a full scan of $\mathbf{T}$ to compute  
$\text{FPR}_\mathbf{T}(h)$ and $\text{Cov}_\mathbf{T}(h)$ for each
hypothesis $h$.
This is both costly and slow
considering the fact that  $\mathbf{T}$ is typically large 
in enterprise data lakes (e.g., in terabytes), 
where a full-scan can take hours. 
This is especially problematic for user-in-the-loop 
validation like in Google's TFDV, where users
are expected to review/modify suggested validation rules
and interactive response time is critical.

We thus introduce an offline index with summary
statistics of $\mathbf{T}$,  so that at online time and given a new 
query column $C$, 
we can evaluate hypothesis patterns $h$ efficiently using only the index, 
without needing to scan $\mathbf{T}$ again.

\begin{figure}[t]
        \centering
        \includegraphics[width=0.5\textwidth]{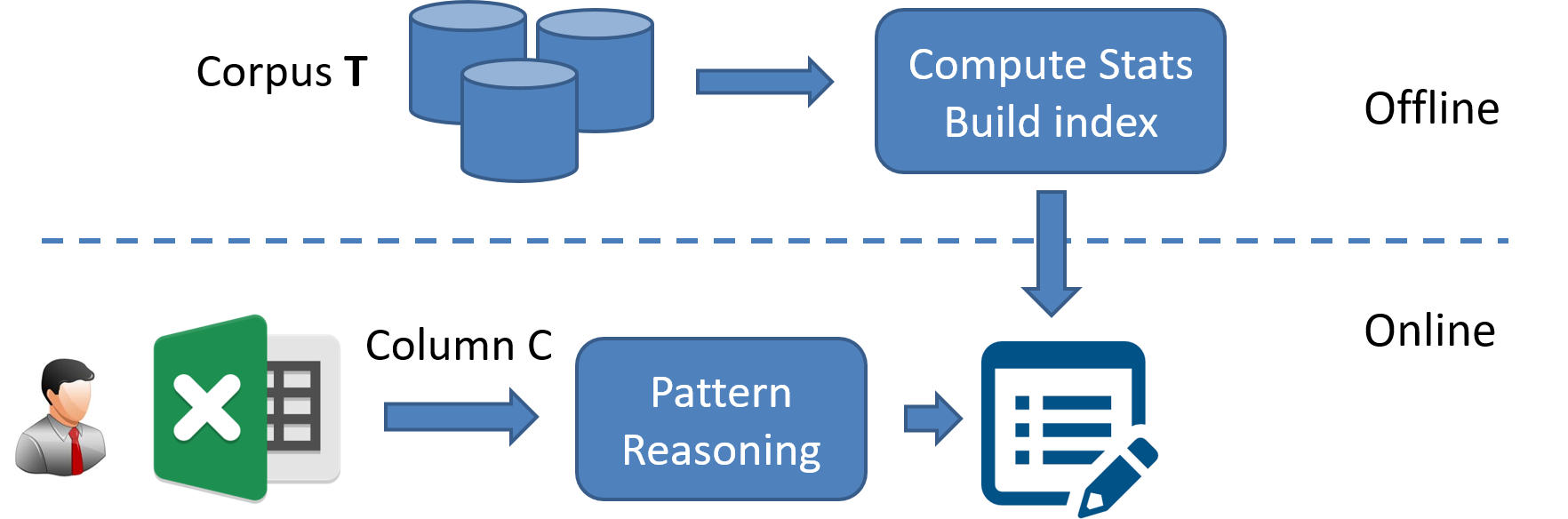}
\caption{Architecture of the \sj{} System.}
\label{fig:architecture}
\end{figure}

The architecture of our system can be
seen in Figure~\ref{fig:architecture}.

\textbf{Offline stage.}
In the offline stage, we perform
one full scan of $\mathbf{T}$, enumerating all possible patterns for
each $D \in \mathbf{T}$ as 
$\mathbf{P}(D) = \cup_{v \in D} \mathbf{P}(v)$,
or the union of patterns for all $v \in D$
(where $\mathbf{P}$ is the generalization hierarchy 
in Section~\ref{sec:language}).

Note that the space of $\mathbf{P}(D)$ 
can be quite large, especially for
``wide'' columns with many tokens as discussed in Figure~\ref{fig:generalization}. 
To limit $\mathbf{P}(D)$, in the basic version
we produce $\mathbf{P}(D)$ as 
$\cup_{v \in D, t(v)<\tau} \mathbf{P}(v)$, where $t(v)$ is the number
of tokens in $v$ (defined as the number of consecutive sequences
of letters, digits, or symbols in $v$), 
and $\tau$ is some fixed constant (e.g., 8 or 13) to make $\mathbf{P}(D)$
tractable.  We will describe how 
columns wider than $\tau$ can be safely skipped in offline 
indexing without affecting result quality, using 
a vertical-cut technique (in Section~\ref{sec:vertical-cuts}).

For each pattern $p \in \mathbf{P}(D)$,
we pre-compute the local impurity score of $p$ on 
$D$ as $\text{Imp}_{D}(p)$,
since $p$ may be hypothesized 
as a domain pattern for some query column $C$ in the future,
for which this $D$ can provide a piece of local evidence.

Let $\mathbf{P}(\mathbf{T}) = \{p | p \in \mathbf{P}(D), D \in \mathbf{T}\}$
be the space of possible patterns in $\mathbf{T}$.
We pre-compute $\text{FPR}_{\mathbf{T}}(p)$ for
all possible $p \in \mathbf{P}(\mathbf{T})$
on the entire $\mathbf{T}$, 
by aggregating local $\text{Imp}_{D}(p)$ scores using
Equation~\eqref{eqn:fpr} and~\eqref{eqn:fpr_agg}.
The coverage scores $\text{Cov}_{\mathbf{T}}(p)$ of all $p$ can be 
pre-computed similarly as 
$\text{Cov}_{\mathbf{T}}(p) = |\{ D | D \in \mathbf{T}, v \in D, p \in \mathbf{P}(v) \}|$.

The result from the offline step is an index for lookup that maps
each possible $p \in \mathbf{P}(\mathbf{T})$, to its
pre-computed $\text{FPR}_{\mathbf{T}}(p)$ 
and $\text{Cov}_{\mathbf{T}}(p)$ values.
We note that this index is many
orders of magnitude smaller than the original
$\mathbf{T}$ -- e.g., a 1TB corpus $\mathbf{T}$ 
yields an index less than 1GB in our experiments.

\textbf{Online stage.}
At online query time, for a given
query column $C$, we enumerate $h \in \mathbf{H}(C)$,
and use the offline index to perform a lookup to retrieve 
$\text{FPR}_{\mathbf{T}}(p)$ and $\text{Cov}_{\mathbf{T}}(p)$
scores for FMDV, without needing to scan the full $\mathbf{T}$ again.
This indexing approach reduces our latency per 
query-column $C$ from hours to
tens of milliseconds, which makes it possible
to do interactive, human-in-the-loop
verification of suggested rules,
as we will report in experiments.

\vspace{-2mm}
\section{Auto-Validate (Vertical Cuts)}
\label{sec:vertical-cuts}

So far we only discussed the basic-version of
\sj. We now describe formulations that handle
more complex and real situations: (1) \textit{composite} structures
in query columns (this section); 
and (2) ``dirty'' values in query columns that violate
the homogeneity assumption (Section~\ref{sec:horizontal-cuts}).

\begin{figure}[t]
        \centering
        \includegraphics[width=0.45\textwidth]{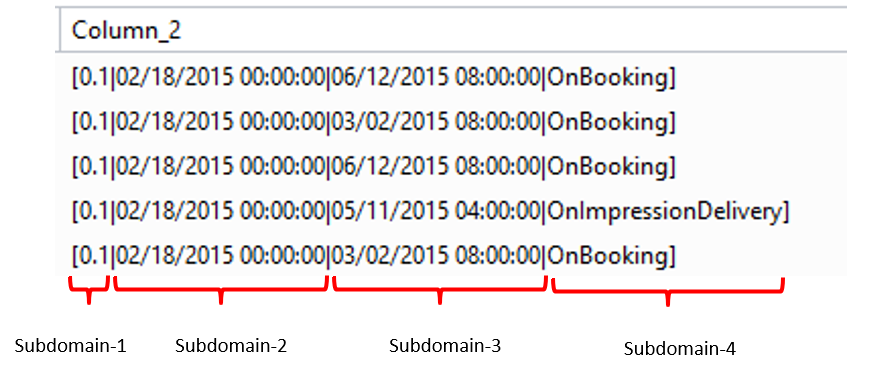}
\caption{Example columns with composite domains.}
\label{fig:example-split}
\end{figure}

Figure~\ref{fig:example-split} shows a real
column with composite structures 
that are concatenated from \textit{atomic domains}. Intuitively, 
we can see that it consists of at least 4
``sub-domains'', a floating-number (``\val{0.1}''),
followed by two time-stamps, 
followed by a status message (``\val{OnBooking}'').
We observe that such columns are common
in machine-generated data, where 
many are complex with over 10 sub-domains.

Such complex columns pose both quality and
scalability challenges. In terms of quality,
if we were to directly test complex
patterns against $\mathbf{T}$ like in FMDV, we may not be
able to see many exact matches in $\mathbf{T}$, 
because the complex patterns in $C$ may be concatenated 
in ad-hoc ways and are
sparsely represented in $\mathbf{T}$.
As a result, we may fail to find a
feasible solution to FMDV, 
because of the coverage requirement in 
Equation~\eqref{eqn:fmdv_cov}.

The second challenge is scalability.
As column $C$ become ``wider'' with
more tokens/segments, the space of possible patterns $\mathbf{P}(C)$
 that we will have to enumerate
grows exponentially in the number of tokens.
This is an issue both at online query time
for query column $C$, as well as
at offline indexing time to enumerate 
$\mathbf{P}(D)$ for data column $D \in \mathbf{T}$ --
recall that the date-time example in 
Figure~\ref{fig:generalization} already produces
over 3 billion patterns;  for the column in
Figure~\ref{fig:example-split} the
space of patterns is simply impractically large to enumerate. 
In the offline indexing step in Section~\ref{sec:architecture},
we discussed that we intentionally omit wide columns with
more than $\tau$ tokens to make pattern enumeration
tractable. We will show here that our
horizontal-cut algorithm
can ``compensate'' such omission without affecting result quality.

A natural approach to complex query column $C$
is to vertically ``cut'' it into
sub-columns (like shown in  
Figure~\ref{fig:example-split}). Patterns for
each sub-column can then be generated in turn using FMDV.
The benefit of this divide-and-conquer
approach is that each sub-domain
is likely well-represented in $\mathbf{T}$,
allowing them to be reliably inferred from $\mathbf{T}$.
Furthermore, the cost of pattern-enumeration in offline-indexing 
becomes significantly smaller, as each sub-domain can be enumerated
separately.

Specifically, we first use a lexer to tokenize each $v \in C$
into coarse-grained token-class (\val{<symbol>}, 
\val{<num>}, \val{<letter>}), by scanning each $v$
from left to right and ``growing'' each token until 
a character of a different class is encountered. 
In the example of Figure~\ref{fig:example-split},
for each row, we would generate an identical token-sequence
``\val{[<num>|<num>/<num>/<num> <num>:<num>:<num>|} \ldots \val{|<letter>]}''.

We then perform multi-sequence alignment 
(MSA)~\cite{carrillo1988multiple}
for all token sequences, before actually performing vertical-cuts.
Recall that the goal of MSA is to find optimal 
alignment across all input sequences 
(using objective functions such as 
sum-of-pair scores~\cite{just2001computational}).
Since MSA is NP-hard in 
general~\cite{just2001computational}, we follow
a standard approach to greedily align one additional sequence at a time.
We note that for homogeneous machine-generated 
data, this often solves MSA optimally.

\begin{example}
\label{ex:align}
For the column in Figure~\ref{fig:example-split},
each value $v \in C$ has an identical token-sequence
with 29 tokens:
``\val{[<num>|<num>/<num>/<num> 
<num>:<num>:<num>|} \ldots \val{|<letter>]}'', and
the aligned sequence using MSA can be
produced trivially as just a sequence with these 29 tokens (no gaps).
\end{example}

After alignment, characters in each $v \in C$ would map to
the aligned sequence of length $n$
(e.g., ``\val{0.1}'' maps to the first \val{<num>} in the aligned
token sequence,
``\val{02}'' maps to the second \val{<num>}, etc.).
Recall that our goal is to vertically split the $n$ tokens 
into $m$ segments so that values mapped to
each segment would correspond to a sub-domain as determined by FMDV.
We define an $m$-segmentation of $C$ as follows.

\begin{definition}
\label{def:segmentation}
Given a column $C$ with $n$ aligned tokens,
define $C[s, e]$ as a segment of $C$ that starts from token position 
$s$ and ends at position $e$, with $s, e \in [n]$.
An $m$-segmentation of $C$, 
is a list of $m$ such segments that
collectively cover $C$, defined as:
$(C[s_1, e_1], C[s_2, e_2],$ $\ldots$ $C[s_m, e_m])$,
where $s_1 = 1, e_m = n$, and $s_{i+1} = e_i + 1~\forall i \in [m]$.
We write $C[s_i, e_i]$ as $C_i$ for short,
and an $m$-segmentation as 
$(C_i : i \in [m])$ for short. 
\end{definition}

Note that each $C_i$ can be seen just like a regular column
``cut'' from the original $C$, with which we can invoke 
the FMDV.
Our overall objective is to jointly find: (1) an $m$-segmentation 
$(C_i : i \in [m])$
from all possible segmentation, denoted by $\mathbf{S}(C)$;
and  (2) furthermore find appropriate patterns 
$h_i \in \mathbf{H}(C_i)$ for each segment
$C_i$, such that the entire $C$ can be validated with a low
expected FPR. Also note that because we limit pattern enumeration
in offline-indexing to columns of at most $\tau$ tokens,
we would not find candidate patterns longer than $\tau$ in offline-index,
and can thus naturally limit the span of each segment to at most $\tau$
($e_i - s_i < \tau$, $\forall i$).
We can write this as an optimization problem FMDV-V defined as follows:

\begin{align}
 \text{(FMDV-V)}~ \min_{\substack{ h_1 \dots h_m \\ h_i \in \mathbf{H}(C_i) \\ (C_i: i \in [m]) \in \mathbf{S}(C)}}~~~& \sum_{h_i} \text{FPR}_\mathbf{T}(h_i)   \label{eqn:fmdvv_obj} \\ 
 \mbox{s.t.} ~~
& \sum_{h_i} \text{FPR}_\mathbf{T}(h_i) \leq r \label{eqn:fmdvv_fpr} \\
 & \text{Cov}_\mathbf{T}(h_i) \geq m,~\forall h_i \label{eqn:fmdvv_cov}
\end{align}

In FMDV-V, we are optimizing over all possible segmentation
$(C_i : i \in [m]) \in \mathbf{S}(C)$, and for each $C_i$
all possible hypothesis patterns $h_i \in \mathbf{H}(C_i)$.
Our objective in Equation~\eqref{eqn:fmdvv_obj} is 
to minimize the sum of 
$\text{FPR}_\mathbf{T}(h_i)$ for all $h_i$.
We should note that this corresponds to a ``pessimistic'' 
approach that assumes non-conforming rows 
of each $h_i$ are disjoint at a row-level, thus needing
to sum them up in the objective function.
(An alternative could ``optimistically'' assume
non-conforming values in different segments 
to overlap at a row-level, in which case
we can use \textit{max} in place of the \textit{sum}.
We find this to be less effective and omit details in the interest of space).


We can show that the minimum FRP scores of 
$m$-segmentation optimized in Equation~\eqref{eqn:fmdvv_obj} have 
\textit{optimal substructure}~\cite{cormen2009introduction}, which makes
it amenable to dynamic programming (without exhaustive
enumeration). Specifically, 
we can show that the following holds:

\begin{equation}
\label{eqn:substructure}
\begin{split}
\text{minFRP}&(C[s, e])  = \\ \min \bigg(&  \min_{h \in \mathbf{P}(C[s, e])} \text{FPR}_{\mathbf{T}}(h), \\ 
&  \min_{t \in[s, e-1]}{\Big( \text{minFPR}(C[s, t]) + \text{minFPR}(C[t+1, e]) \Big)}\bigg)
\end{split}
\end{equation}

Here, minFRP$(C[s, e])$ is the minimum FRP score with
vertical splits as optimized in 
Equation~\eqref{eqn:fmdvv_obj}, for a sub-column
corresponding to the segment
$C[s, e]$ from token-position $s$ to $e$.
Note that $C[1, n]$ corresponds to the original query column $C$,
and minFRP$(C[1, n])$ is the score we want to optimize.
Equation~\eqref{eqn:substructure} shows how this
can be computed incrementally.
Specifically, minFRP$(C[s, e])$ is the minimum of
(1) $\min_{h \in \mathbf{P}(C[s, e])}\text{FPR}_{\mathbf{T}}(h)$,
which is the FRP score 
of treating $C[s, e]$ as one column (with no splits), solvable
using FMDV;
and (2) $\min_{ t \in [s, e-1]}{\big( \text{minFPR}(C[s, t]) + \text{minFPR}(C[t+1, e]) \big)}$, which is the best minFRP scores from 
all possible two-way splits that
can be solved optimally in polynomial time using bottom-up 
dynamic-programming.

\begin{example}
\label{ex:fmdv-v}
Continue with Example~\ref{ex:align}.
The aligned sequence of column $C$ in Figure~\ref{fig:example-split}
has 29 tokens, corresponding to $C[1, 29]$. 
The sub-sequence $C[4, 14]$ maps to identical
values ``\val{02/18/2015 00:00:00}'' for all rows, 
which can be seen as a ``sub-column'' split from $C$.

Using Equation~\eqref{eqn:substructure}, 
\text{minFRP}$(C[4, 14])$ of this sub-column 
can be computed as 
the minimum of:
(1) Not splitting of $C[4, 14]$, or
$\min_{h \in \mathbf{P}(C[4, 14])}\text{FPR}_{\mathbf{T}}(h)$,
computed using FMDV; and
(2) Further splitting $C[4, 14]$ into two parts, 
where the best score is computed as
$\min_{t \in[4, 13]}{\Big( \text{minFPR}(C[4, t]) + \text{minFPR}(C[t+1, 14]) \Big)}$, which can in turn be recursively computed.
Suppose we compute (1) to have FPR of 0.004\%, 
and the best of (2) to have FPR of 0.01\% (say, splitting 
into ``\val{02/18/2015}'' and ``\val{00:00:00}'').
Overall, not splitting $C[4, 14]$ is the best option.

Each such $C[s, e]$ can be computed bottom-up, until reaching
the top-level. 
The resulting split
minimizes the overall FPR and is the solution to FMDV-V.
\end{example}

We would like to highlight that by using vertical-cuts, we can safely
ignore wide columns with more than $\tau$ tokens during offline
indexing (Section~\ref{sec:architecture}), without
affecting result quality. As a concrete example, 
when we encounter a column like
in Figure~\ref{fig:example-split}, brute-force enumeration would
generate at least $4^{29}$ patterns, which is impractically large
and thus omitted given a smaller token-length limit $\tau$ (e.g., 13) 
in offline indexing. However, when we encounter 
this column Figure~\ref{fig:example-split}
as a query column, we can still generate valid domain patterns
with the help of vertical-cuts (like shown in
Example~\ref{ex:fmdv-v}),
as long as each segment has at most
$\tau$ tokens. Vertical-cuts thus
greatly speed up offline-indexing as
well as online-inference, without affecting result quality.

\section{Auto-Validate (Horizontal Cuts)}
\label{sec:horizontal-cuts}
\begin{figure}
        \centering
        \includegraphics[width=0.5\textwidth]{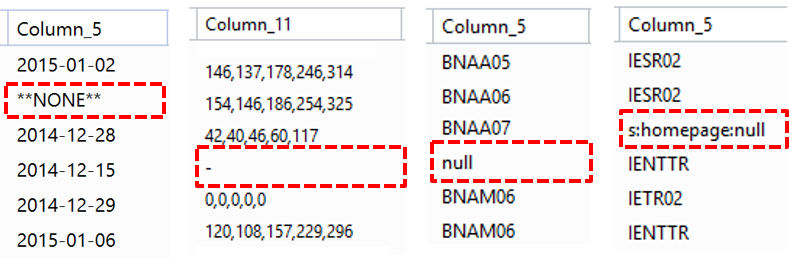}
\caption{Example ad-hoc special values (in red boxes)
not observing patterns of other values in the column.}
\label{fig:example-nulls}
\end{figure}

So far we assumed the query column $C$
to be ``clean'' with homogeneous
values drawn from the same domain 
(since they are generated by the same underlying program).
While this is generally true, we observe that some
columns can have ad-hoc special values  
(e.g., denoting null/empty-values) that do
not follow the domain-pattern of $C$,
as shown in Figure~\ref{fig:example-nulls}.

We note that examples like these are not uncommon -- even for
machine-generated data, there can be a 
branch of logic (think of \val{try-except}) that
 handles special cases (e.g., nulls),
and produces ad-hoc values 
not following the pattern of ``normal'' values.
We henceforth refer to these as \textit{non-conforming values}.

Recall that in FMDV, by assuming that the query column $C$ has
homogeneous values, we select patterns
from $\mathbf{H}(C) = \cap_{v \in C}{\mathbf{P}(v)}$, 
or the intersection of patterns for all values  $v \in C$.
Such an assumption does not hold 
for columns in Figure~\ref{fig:example-nulls},
yielding an empty $\mathbf{H}(C)$ and
no feasible solution to FMDV.

We thus consider a variant of \sj with ``horizontal
cuts'', meaning that we can ``cut off'' a small fraction
of non-conforming values in 
$C$ (which is identical to making patterns tolerate
a small fraction of non-conforming values). 
We use a \textit{tolerance parameter}
$\theta$ to control the maximum allowed fraction of 
non-conforming values that can be cut (e.g., 1\%, 5\%, etc.), 
which allows for a conscious trade-off between 
precision and recall.

This optimization problem, termed
FMDV-H, is defined as

\begin{align}
\hspace{-1cm} \text{(FMDV-H)}~~\min_{h}~& \text{FPR}_\mathbf{T}(h)   \label{eqn:fmdvh_obj} \\ 
 \mbox{s.t.} ~~ & h \in \cup_{v\in C}{\mathbf{P}(v) \setminus \text{``.*''}} \label{eqn:fmdvh_h} \\
&  \text{FPR}_\mathbf{T}(h) \leq r \label{eqn:fmdvh_fpr} \\
 & \text{Cov}_\mathbf{T}(h) \geq m \label{eqn:fmdvh_cov} \\
 & |\{ v | v \in C, h \in \mathbf{P}(v) \}| \geq (1-\theta) |C| \label{eqn:fmdvh_tolerance}
\end{align}

Like before, our objective function is to 
minimize FPR in Equation~\eqref{eqn:fmdvh_obj}.
Equation~\eqref{eqn:fmdvh_h} shows that the 
hypothesis pattern $h$ is drawn from
the union of possible patterns for all $v \in C$.
The FPR and Coverage requirements in
Equation~\eqref{eqn:fmdvh_fpr} and
Equation~\eqref{eqn:fmdvh_cov} are 
the same as the basic FMDV.
Finally, Equation~\eqref{eqn:fmdvh_tolerance} requires
that the selected $h$ has to match at least 
$(1-\theta)$ fraction of values in $C$, where $\theta$ 
is the tolerance parameter above (the remaining
 $\theta$ fraction are non-conforming values).

\begin{example}
\label{ex:fmdvh}
Consider the leftmost column $C$ in Figure~\ref{fig:example-nulls}.
Using FMDV-H, we determine the pattern $h=$
``\val{<digit>+,<digit>+,<digit>+,} \val{<digit>+,<digit>+}''
to be a valid solution, as it meets the FPR and coverage
requirements based on $\mathbf{T}$.
Furthermore $h$ is consistent with 99\% of values in $C$
(the remaining 1\% non-conforming value
is the ``\val{-}'' marked in a red box), thus also 
satisfying Equation~\eqref{eqn:fmdvh_tolerance}.
\end{example}

For arbitrary $\theta$ and generalization hierarchy,
we can show that deciding whether there is a feasible 
solution to FMDV-H is NP-hard (let alone minimizing FPR), using a reduction
from independent set~\cite{karp1985fast}.

\begin{theorem}
\label{thm:fmdvh}
The decision version of FMDV-H is NP-hard.
\end{theorem}

While FMDV-H is hard in general, in practice because
the patterns of non-conforming values often
do not intersect with those of ``normal'' values
(as in Example~\ref{ex:fmdvh}),
they create 
easier instances of FMDV-H.
Leveraging this observation, in this work 
we optimize FMDV-H greedily, by discarding
values whose patterns do not intersect
with most others in $C$.
We can then find the optimal pattern for the
remaining conforming values in FMDV-H 
using FMDV.

\textbf{Distributional test of non-conforming values.}
Given a pattern $h$ inferred from the
``training data'' $C$ using FMDV-H, and given the
data $C'$ arriving in the future, our last task is 
to determine whether the fraction of non-conforming
values in $C'$ has changed significantly from $C$.

Specifically, at ``training'' time,
we can calculate the fraction of values in $C$ not 
conforming to $h$, denoted as $\theta_{C}(h) 
= \frac{|\{ v | v \in C, h \notin \mathbf{P}(v) \}|}{|C|}$.
At ``testing'' time, we also compute the fraction of 
values in $C'$ not 
matching $h$, denoted as $\theta_{C'}(h)$.

A naive approach to validate $C'$ is to trigger alarms
if $\theta_{C'}(h)$ is greater than $\theta_{C}(h)$.
This, however, is prone to false-positives.
Imagine a scenario where we compute the non-conforming ratio
$\theta_{C}(h)$ on training data $C$ to be 0.1\%. 
Suppose on $C'$ we find $\theta_{C'}(h)$ to be 0.11\%.
Raising alarms would likely be false-positives.
However, if $\theta_{C'}(h)$ is substantially higher, say at 
5\%, intuitively it becomes an issue
that we should report. (The special
case where no value in $C'$ matches $h$
has the extreme $\theta_{C'}(h)=$ 100\%).

To reason about it formally, we 
model the process of drawing a conforming value
vs. a non-conforming value in
$C$ and $C'$, as sampling from two binomial distributions.
We then perform a form of statistical hypothesis test called 
\textit{two-sample homogeneity test},
to determine whether the fraction of non-conforming values
has changed significantly,
using $\theta_{C}(h)$, $\theta_{C'}(h)$,
and their respective sample sizes $|C|$ and $|C'|$.

Here, the null-hypothesis $H_0$ states that the
two underlying binomial distributions are the same;
whereas $H_1$ states that the reverse is true. 
We use \textit{Fischer's exact test}~\cite{agresti1992survey}
and \textit{Pearson's Chi-Squared test}~\cite{kanji2006} with
Yates correction for this purpose, both of which are suitable tests
in this setting.
We report $C'$ as an issue if the divergence from $C$ 
is so strong such that the null hypothesis can be rejected
(e.g., $\theta_{C}(h) =$ 0.1\% and
$\theta_{C'}(h) =$ 5\%).

We omit details of these two
statistical tests in the interest of space. 
In practice we find both to perform well, with 
little difference in terms of validation quality.

\iftoggle{fullversion}{
\textbf{Time Complexity.}
Overall, the time complexity of the offline step is 
$\mathcal{O}(|\mathbf{T}| |\mathbf{P}(D)|)$,
which comes from scanning each column $D$ in
$\mathbf{T}$ and enumerating its possible patterns. 
Recall that because we use vertical-cuts to handle
composite domains, $\mathbf{P}(D)$ can be constrained 
by the chosen constant $\tau$ that is the upper-limit
of token-count considered in the indexing step (described
in Section~\ref{sec:architecture}), without affecting result
quality because at online inference time we compose domains
using vertical-cuts
(Section~\ref{sec:vertical-cuts}).

In the online step, the complexity of the most involved
FMDV-VH variant is $\mathcal{O}(|\mathbf{P}(v)| |C| t(C)^2)$,
where $t(C)$ is the max number of tokens in query column $C$.
Empirically we find the overall latency to be less than 100ms on average,
as we will report in our experiments.
}



\vspace{-2mm}
\section{Experiments}
We implement our offline indexing algorithm in a Map-Reduce-like system
used internally at Microsoft~\cite{chaiken2008scope, zhou2012scope}. 
The offline job processes 7M columns with over 1TB data in under 3 hours. 

\vspace{-2mm}
\subsection{Benchmark Evaluation}
We build benchmarks using real data, to evaluate the quality of data-validation rules generated 
by different algorithms.

\textbf{Data set}. We evaluate algorithms using 
two real corpora:  

$\bullet$ \texttt{Enterprise:} We crawled data produced by
production pipelines from an enterprise data lake
at Microsoft. Over 100K daily production 
pipelines (for Bing, Office, etc.)-eps-converted-to.pdf
read/write data in the lake~\cite{zhou2012scope}.

$\bullet$ \texttt{Government}: We crawled data files
in the health domain from NationalArchives.gov.uk, following a procedure suggested in~\cite{bogatu2020dataset} 
(e.g., using queries like ``hospitals''). We used the crawler
from the authors of~\cite{bogatu2020dataset} (code is available from~\cite{crawler}).  These data files correspond to data in a different
domain (government).\footnote{This government benchmark is released at \url{https://github.com/jiesongk/auto-validate}.}

We refer to these two corpora as $\mathbf{T_{E}}$ and
$\mathbf{T_{G}}$, respectively.
The statistics of the two corpora can be found in f~\ref{tab:corpora}.

\begin{table}[h]
\centering
\scriptsize
\begin{tabular}{|c | c | c | c | c | } 
 \hline
 Corpus & \begin{tabular}{@{}c@{}}total \# of \\ data files\end{tabular} & \begin{tabular}{@{}c@{}}total \# of \\ data cols\end{tabular} & \begin{tabular}{@{}c@{}} avg col value cnt \\ (standard deviation)\end{tabular}   & \begin{tabular}{@{}c@{}} avg col distinct value cnt \\ (standard deviation) \end{tabular}  \\  \hline
  \texttt{Enterprise} ($\mathbf{T_{E}}$) & 507K & 7.2M & 8945 (17778) & 1543 (7219) \\  \hline
 \texttt{Government} ($\mathbf{T_{G}}$) & 29K & 628K & 305 (331) &  46 (119) \\
 \hline
\end{tabular}
\caption{Characteristics of data corpora used.}
\label{tab:corpora}
\end{table}

\textbf{Evaluation methodology}.  
We randomly sample 1000 columns
from $\mathbf{T_{E}}$ and $\mathbf{T_{G}}$, respectively, 
to produce two benchmark sets of query-columns, denoted by
$\mathbf{B_{E}}$ and $\mathbf{B_{G}}$.\footnote{We use 
the first 1000 values of each column in $\mathbf{B_{E}}$
and the first 100 values of each column in 
$\mathbf{B_{G}}$ to control 
column size variations.}

Given a benchmark $\mathbf{B}$ with 1000 columns, 
$\mathbf{B} = \{C_i | i \in [1, 1000]\}$,
we programmatically 
evaluate the precision and
recall of data-validation rules generated on $\mathbf{B}$ 
as follows. For each column $C_i \in \mathbf{B}$, 
we use the first $10\%$ of values in $C_i$
as the ``training data'' that arrive first, 
denoted by $C_i^\text{train}$, from which
validation-rules need to be inferred. The remaining
$90\%$ of values are used as ``testing data''  $C_i^\text{test}$,
which arrive in the future but will be tested against the inferred
validation-rules.

Each algorithm $A$ can observe $C_i^\text{train}$ 
and ``learn'' data-validation 
rule $A(C_i^\text{train})$.
The inferred rule $A(C_i^\text{train})$ will be ``tested'' on two groups of
``testing data'': (1) $C_i^\text{test}$; and (2) $\{C_j | C_j \in \mathbf{B}, j \neq i\}$.

For (1), we know that when $A(C_i^\text{train})$ is used to 
validate against $C_i^\text{test}$, 
no errors should be detected
because $C_i^\text{test}$ and $C_i^\text{train}$ were 
drawn from the same column $C_i \in \mathbf{B}$.
If it so happens that $A(C_i^\text{train})$ incorrectly reports any error
on $C_i^\text{test}$, it is likely a false-positive.

For (2), when $A(C_i^\text{train})$ is used to validate 
$\{C_j | C_j \in \mathbf{B}, j \neq i\}$,
we know each $C_j$ ($j \neq i$) is likely from a different 
domain as $C_i$, and should be 
flagged by $A(C_i^\text{train})$. (This simulates schema-drift
and schema-alignment errors that are
common, because upstream data can add or remove
columns). This allows us to
evaluate the ``recall'' of each $A(C_i^\text{train})$.

More formally, we define the \textit{precision} of 
algorithm $A$ on test case $C_i$,
denoted by $P_A(C_i)$, as 1 if no value in $C_i^{test}$
 is incorrectly flagged
by $A(C_i^\text{train})$ as an error, and 0 otherwise.
The overall precision on benchmark $\mathbf{B}$ 
is the average across all cases:
$P_{A}(\mathbf{B})  = \avg_{C_i \in \mathbf{B}} P_A(C_i)$.

The \textit{recall} of algorithm $A$ on case $C_i$, 
denoted by $R_A(C_i)$, is defined as the fraction of 
cases in $\{C_j | C_j \in \mathbf{B}, j \neq i\}$
that $A(C_i^\text{train})$
can correctly report as errors:
\begin{equation}
R_{A}(C_i)  = \frac{|\{C_j |\text{Validate}(C_j, A(C_i^\text{train})) = \text{fail}, C_j \in \mathbf{B}, j \neq i \}|}{|\{C_j | C_j \in \mathbf{B}, j \neq i \}|}
\end{equation}
Since high precision is critical for data-validation,
if algorithm $A$ produces false-alarms on case $C_i$,
we squash its recall $R_{A}(C_i)$  to 0. 
The overall recall across all cases in $\mathbf{B}$ is then:
$R_{A}(\mathbf{B})  = \avg_{C_i \in \mathbf{B}} R_A(C_i)$.

In addition to programmatic evaluation, we also
manually labeled each case in $\mathbf{B_{E}}$
with an ideal ground-truth validation-pattern. We use
these ground-truth labels, to both
accurately report precision/recall on $\mathbf{B_{E}}$,
and to confirm the validity of our programmatic evaluation
methodology.

\vspace{-2mm}
\subsection{Methods Compared}
\label{sec:methodsCompared}
We compare the following algorithms using each
benchmark $\mathbf{B}$, by reporting precision/recall numbers
$P_A(\mathbf{B})$ and $R_{A}(\mathbf{B})$.

\textbf{\sj}. This is our proposed approach. To understand the behavior of different variants of FMDV, we 
report FMDV, FMDV-V, FMDV-H, as well as a variant FMDV-VH
which combines vertical and horizontal cuts. For FMDV-H
and FMDV-VH, we report numbers using two-tailed
Fischer's exact test with a significance level of 0.01.

\textbf{Deequ}~\cite{schelter2019unit}. Deequ is a 
pioneering library from Amazon for declarative 
data validation. It is capable of suggesting data validation
rules based on  training data.
We compare with two relevant rules in Deequ for string-valued data
(version 1.0.2): the
CategoricalRangeRule (refer to as \texttt{Deequ-Cat}) and 
FractionalCategoricalRangeRule (referred to as 
\texttt{Deequ-Fra})~\cite{Deequ},
which learn fixed dictionaries
and require future test data to be in the dictionaries,
either completely (\texttt{Deequ-Cat}) or partially (\texttt{Deequ-Fra}).

\textbf{TensorFlow Data Validation (TFDV)}~\cite{TFDV}. 
TFDV is another pioneering data 
validation library for machine learning pipelines in TensorFlow. It can
also infer dictionary-based validation rules (similar to  \texttt{Deequ-Cat}).
We install TFDV via Python pip (version 0.15.0) and invoke it directly from Python.

\textbf{Potter's Wheel (PWheel)}~\cite{raman2001potter}. This is
an influential pattern-profiling method, which finds the best
pattern for a data column based on 
minimal description length (MDL). 

\textbf{SQL Server Integration Services 
(SSIS)}~\cite{ssis-profiling}. SQL Server 
has a data-profiling feature in SSIS. We invoke it programmatically
to produce regex patterns
for each column.

\textbf{XSystem}~\cite{ilyas2018extracting}. This recent
approach develops a flexible branch and
merges strategy to pattern profiling. We use the authors' implementation
on GitHub~\cite{xsystem-code} to produce patterns.

\textbf{FlashProfile}~\cite{padhi2018flashprofile}. FlashProfile
is a recent approach to pattern profiling, which clusters
similar values by a distance score. 
We use the authors' implementation
in NuGet~\cite{FlashProfileCode} to produce 
regex patterns for each column.

\textbf{Grok Patterns (Grok)}~\cite{grok}. Grok has a collection of
60+ manually-curated regex patterns, widely used in log parsing
and by vendors 
like AWS Glue ETL~\cite{Glue}, to
recognize common data-types
(e.g., time-stamp, ip-address, etc.). For data-validation,
we use all values in $C_i^\text{train}$ to determine whether there
is a match with a known Grok pattern (e.g., ip-address). This approach
is likely high-precision but low recall
because only common data-types are curated.

\textbf{Schema-Matching}~\cite{rahm2001survey}. Because \sj{}
leverages related tables in $\mathbf{T}$ as additional evidence to derive
validation patterns, we also compare with vanilla schema-matching  
that ``broaden'' the training examples using related
tables in $\mathbf{T}$. Specifically, we compare
with two instance-based techniques 
\val{Schema-Matching-Instance-1 (SM-I-1)} 
and \val{Schema-Matching-Instance-10 (SM-I-10)},
which use any column in $\mathbf{T}$ that overlaps with
more than 1 or 10 instances of $C_i^\text{train}$, respectively,
as additional training-examples.
We also compare with two pattern-based
\val{Schema-Matching-Pattern-Majority (SM-P-M)} 
and \val{Schema-Matching-Pattern-Plurality (SM-P-P)},
which use as training-data 
columns in $\mathbf{T}$ whose majority-pattern 
and plurality-patterns match those of 
$C_i^\text{train}$, respectively.  We invoke \val{PWheel}
on the resulting data, since it is the best-performing
pattern-profiling technique in our experiments.

\textbf{Functional-dependency-upper-bound (FD-UB)}. 
Functional dependency (FD) is an orthogonal approach that
leverages multi-column dependency for 
data quality~\cite{rahm2000data}. 
While FDs inferred from individual table instances
often may not hold in a semantic 
sense~\cite{berti2018discovery, papenbrock2016hybrid}, 
we nevertheless evaluate the
fraction of benchmark columns that are part of any FD
from their original tables,
which would be a recall upper-bound for FD-based approaches.
For simplicity, in this analysis, we 
assume a perfect precision for FD-based methods.

\textbf{Auto-Detect-upper-bound (AD-UB)}~\cite{huang2018auto}. 
Auto-detect is a recent approach that detects incompatibility 
through two common patterns that are rarely co-occurring.
For a pair of values $(v_1, v_2)$ to be recognized as incompatible
in Auto-detect, both of $v_1$ and $v_2$ need to correspond to
common patterns, which limits its coverage. 
Like in \texttt{FD-UB}, we evaluate the recall 
upper-bound of Auto-detect (assuming a perfect precision).



\vspace{-2mm}
\subsection{Experimental Results}
\begin{figure*}
\label{fig:result_pr}
    \centering
    \subfigure[\texttt{Enterprise} benchmark:
Numbers reported using a subset of 571 cases where syntactic patterns exist.]{\label{fig:result_pr_ent_scaled}\includegraphics[width=0.75\columnwidth]{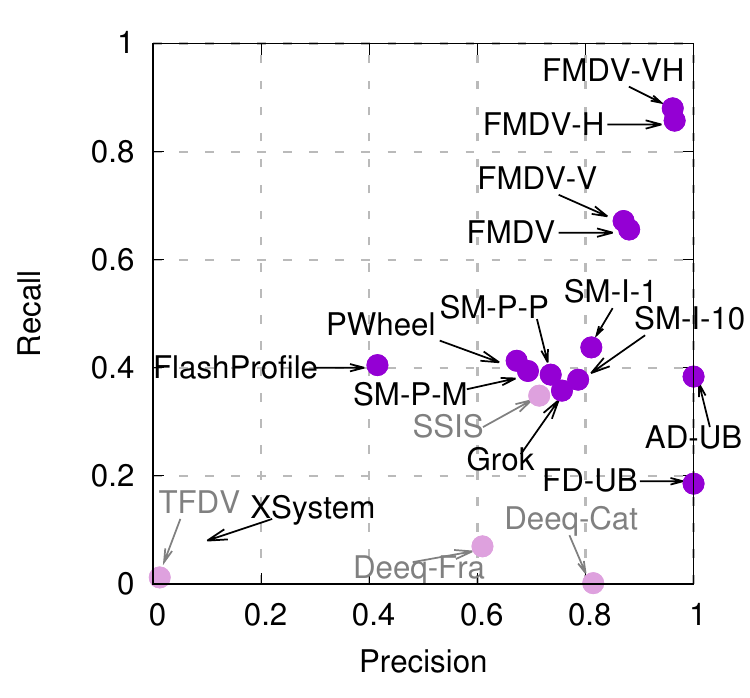}}
    \qquad
    \subfigure[\texttt{Government} benchmark:
Numbers reported using a subset of 359 cases where syntactic patterns exist.
]{\label{fig:result_pr_gov_scaled}\includegraphics[width=0.75\columnwidth]{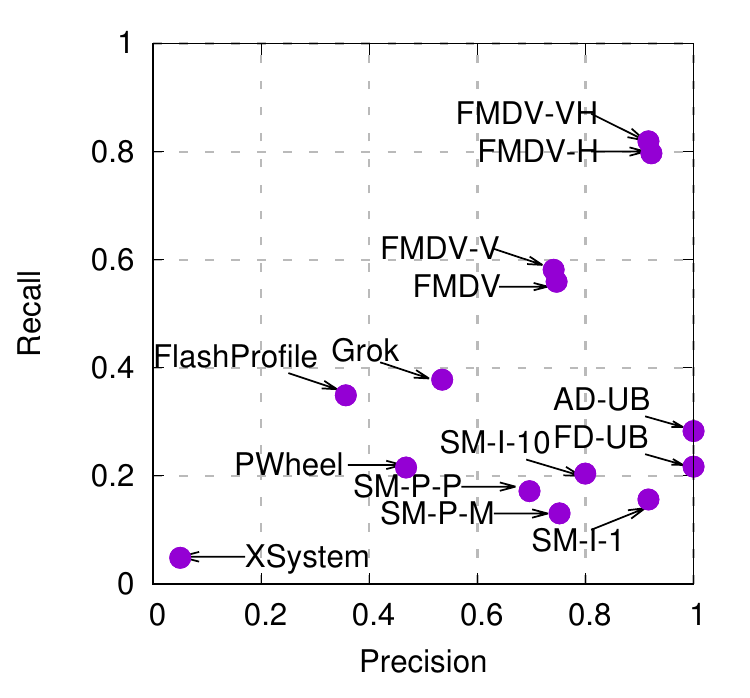}}
    \caption[]{Recall vs. Precision comparisons of all methods, on \texttt{Enterprise} and \texttt{Government} benchmarks. 
}
    \label{fig:accuracy}
\vspace{-3mm}
\end{figure*}

\textbf{Accuracy.}
Figure~\ref{fig:result_pr_ent_scaled} 
shows average precision/recall of all 
methods using the enterprise benchmark $\mathbf{B_E}$ with 
1000 randomly sampled test cases.
Since no pattern-based methods (\val{FMDV}, \val{PWheel}, 
\val{SSIS}, \val{XSystem}, etc.) 
can generate meaningful patterns (except
the trivial ``\val{.*}'') 
on 429 out of the 1000 cases (because they have substantial 
natural language content for example), we report results
on the remaining 571  cases
in Figure~\ref{fig:result_pr_ent_scaled}, where pattern-based 
methods are applicable.

It can be seen from Figure~\ref{fig:result_pr_ent_scaled}
that variants of FMDV are substantially better than other
methods in both precision and recall,
with FMDV-VH being the best at 0.96 precision and 0.88 recall on average.
FMDV-VH is better than FMDV-H,
which is in turn better than 
FMDV-V and FMDV, showing the benefit of using 
vertical-cuts and horizontal-cuts, respectively.

Among all the baselines, we find 
\val{PWheel} and \val{SM-I-1} (which uses schema-matching
with 1-instance overlap)
to be the most competitive, indicating that patterns
are indeed applicable to validating string-valued data, 
but need to be carefully selected to be effective.

Our experiments on \texttt{FD-UB} confirm its orthogonal 
nature to single-column constraints considered by \sj --
the upper-bound recall of \texttt{FD-UB} only ``covers'' around 25\%
of cases handled by \sj 
(assuming discovered FDs have perfect precision).

The two data-validation
methods \texttt{TFDV} and \texttt{Deequ} do not 
perform well on string-valued data, 
partly because their current focus is
on numeric-data, and both use relatively simple 
logic for string data (e.g., dictionaries), which 
often leads to false-positives.

Similar results can be observed on the government benchmark 
$\mathbf{B_G}$,
which is shown in 
Figure~\ref{fig:result_pr_gov_scaled}. (To avoid clutter, we omit 
methods that are not competitive in 
$\mathbf{B_E}$ in this figure).
This benchmark is more challenging because we 
have a smaller corpus and less clean data (e.g.,
many are manually-edited csv files), which leads to lower precision/recall
for all methods. Nevertheless, FMDV methods
still produce the best quality, showing the robustness of
the proposed approaches on challenging data corpus.

\begin{table}[h!]
\vspace{-2mm}
\centering
\small
\begin{tabular}{|c | c| c|} 
 \hline
  Evaluation Method &  precision & recall  \\  \hline
  Programmatic evaluation  &   0.961  &  0.880 \\  \hline 
 Hand curated ground-truth &  0.963  & 0.915 \\  \hline
\end{tabular}
\caption{Quality results on $\mathbf{B_{E}}$, using programmatic evaluation vs. manually-cleaned ground-truth.}
\label{tab:compare-eval}
\end{table}

For all comparisons in Figure~\ref{fig:accuracy}
discussed above, we perform statistical tests using the F-scores
between FMDV-VH and all other methods on 1000 columns 
(where the null-hypothesis being the F-score of 
FMDV-VH is no better than a given baseline). We report
that the $p$-values of all comparisons range from 0.001 to 0.007
(substantially smaller than the $p<0.05$ level),
indicating that we should reject the null-hypothesis and the observed advantages are likely significant.

\textbf{Manually-labeled ground-truth.}
While our programmatic evaluation provides a good proxy of the ground-truth
without needing any labeling effort, we also perform a manual
evaluation to verify the validity of the programmatic evaluation.

Specifically, we manually label the ground-truth validation patterns
for 1000 test cases in $\mathbf{B_{E}}$, 
and perform two
types of adjustments: (1) To accurately report precision, 
we manually remove values in the test-set of each
column that should not belong to the column
(e.g., occasionally column-headers are parsed
incorrectly as data values), to ensure that we do not
unfairly punish methods that identify correct patterns; and 
(2) To accurately report recall, we identify
ground-truth patterns of each test query column, so that in recall
evaluation if another column is drawn from the same domain
with the identical pattern, 
we do not count it as a recall-loss.

We note that both adjustments \textit{improve}
the precision/recall, because our programmatic evaluation
\textit{under-estimates} the true precision/recall. 
Table~\ref{tab:compare-eval} compares the quality results
using programmatic evaluation and manual ground-truth,
which confirms the validity of the programmatic evaluation.

\begin{figure*}
\vspace{4mm}
  \includegraphics[width=0.9\textwidth]{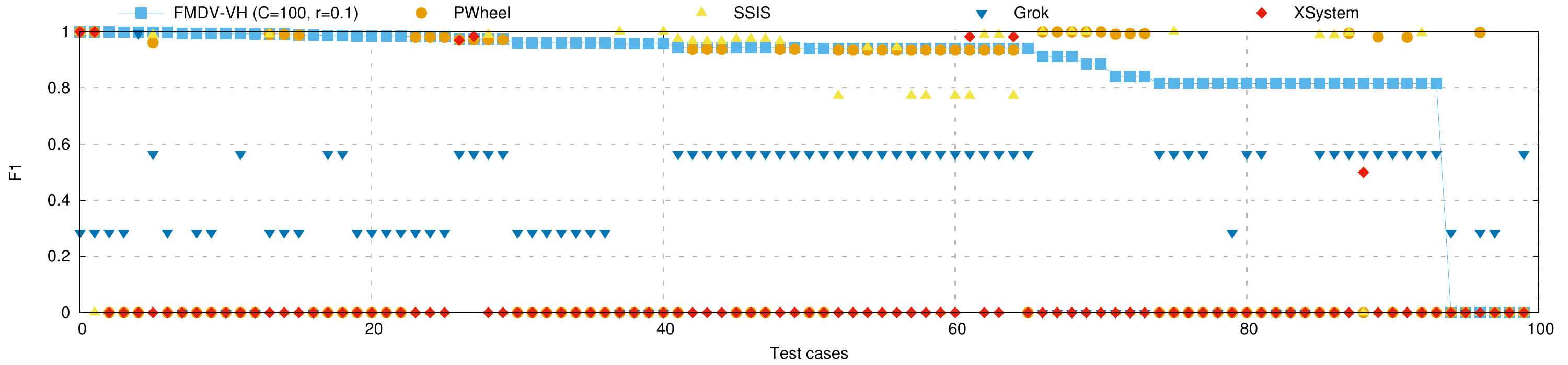}
  \caption{Case-by-case results reported using F-1 scores, on 100 randomly sampled cases.}
  \label{fig:case_by_case}
\end{figure*}

Figure~\ref{fig:case_by_case} shows 
a case-by-case analysis of F1 results
on $\mathbf{B_E}$ with competitive
methods, using 100 randomly sampled
cases, sorted by their results on 
FMDV-VH to make comparison easy.
We can see that FMDV dominates other 
methods. An error-analysis on
the failed cases shows that these are mainly attributable
to advanced pattern-constructs such as 
flexibly-formatted URLs, and unions of distinct patterns, 
which are not supported by our current profiler. 

\begin{figure*}
    \centering
    \subfigure[Sensitivity to FPR threshold ($r$)]{\label{fig:sensitivity_r}\includegraphics[height=5.7cm]{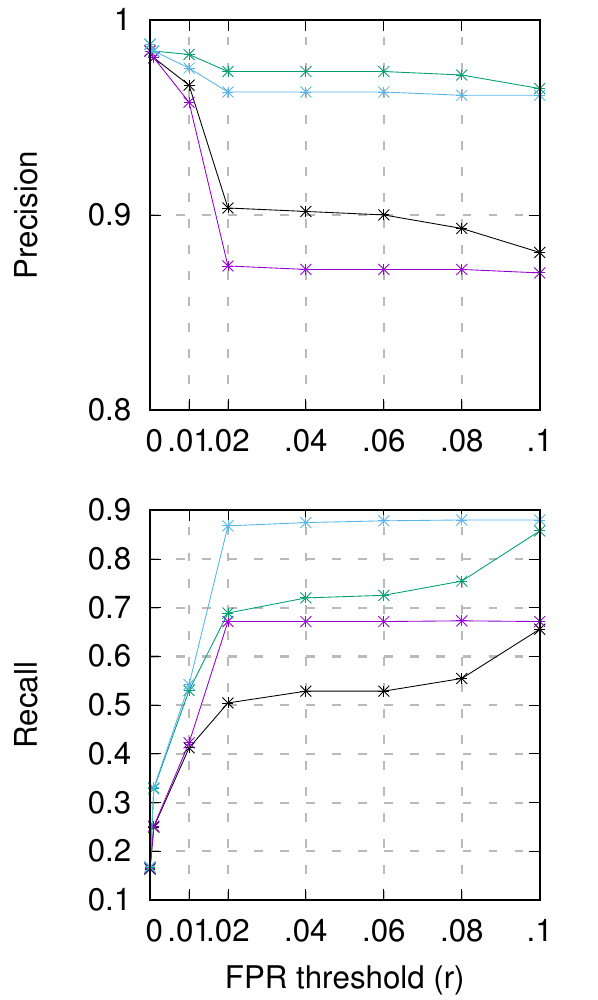}}
    \qquad
    \subfigure[Sensitivity to Coverage ($m$)]{\label{fig:sensitivity_C}\includegraphics[height=5.7cm]{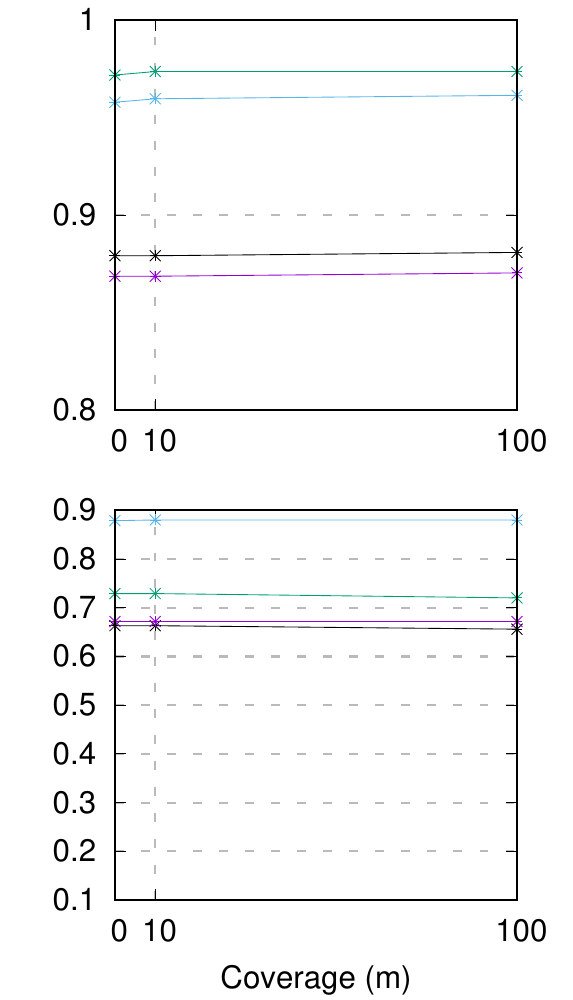}}
    \qquad
    \subfigure[Sensitivity to token-limit ($\tau$)]{\label{fig:sensitivity_to_index_tau}\includegraphics[height=5.7cm]{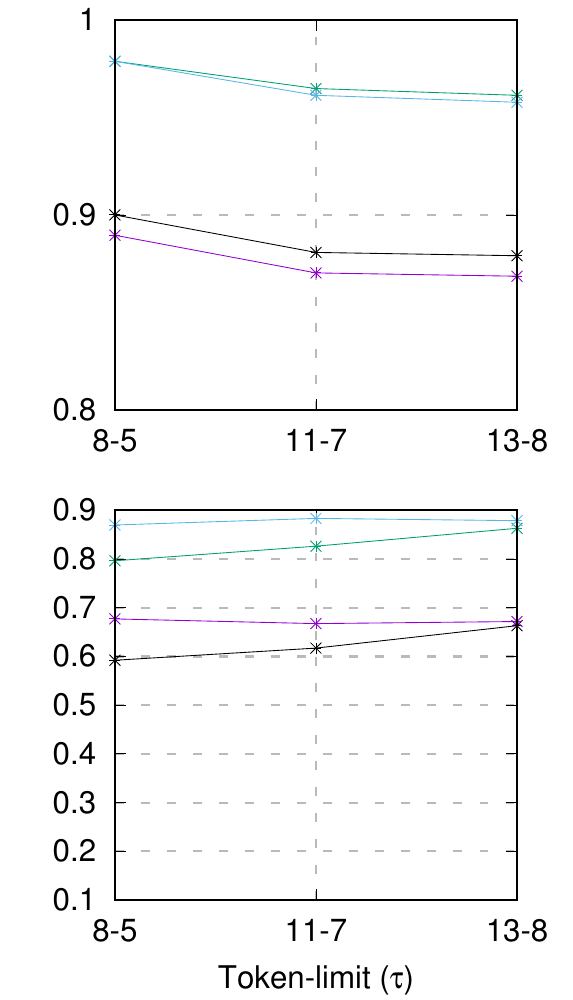}}
    \qquad
    \subfigure[Sensitivity of FMDV-H to ($\theta$)]{\label{fig:sensitivity_to_theta}\includegraphics[height=5.7cm]{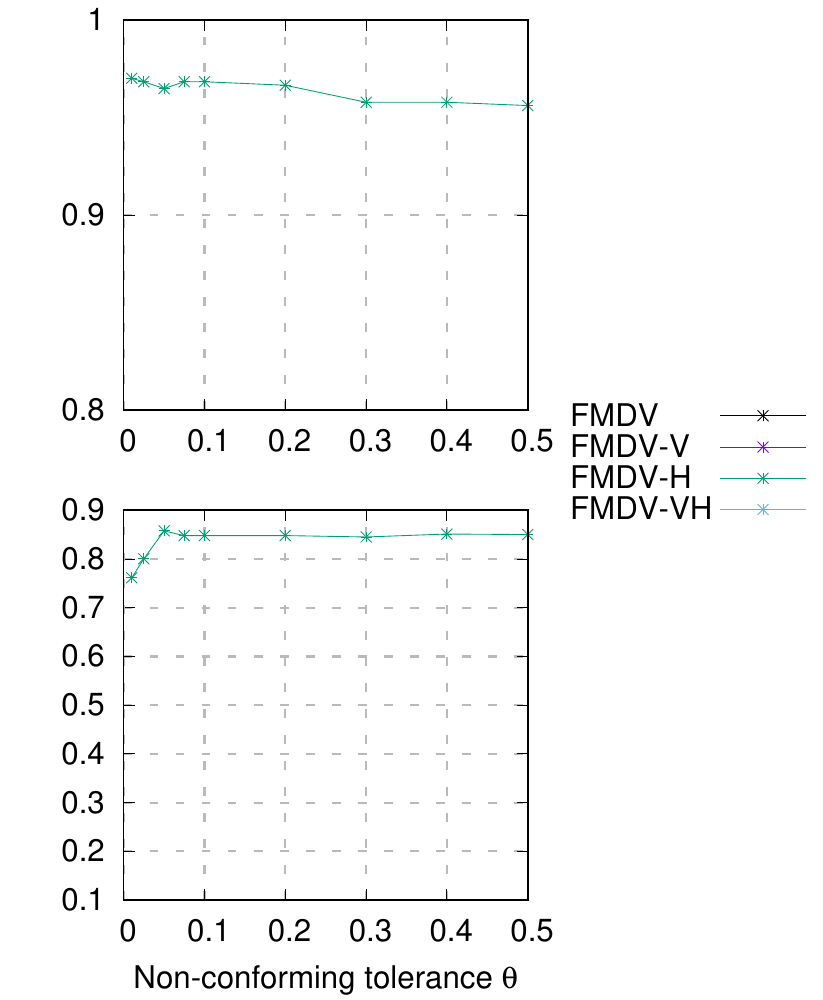}}
    
    \caption[]{Sensitivity analysis of FMDV algorithms on the Enterprise benchmark.}
    \label{fig:sensitivity}
\end{figure*}

\textbf{Sensitivity.}
Figure~\ref{fig:sensitivity} shows a detailed sensitivity analysis for
all FMDV variants using average precision/recall.
Figure~\ref{fig:sensitivity_r} shows the result with a
varying FRP target $r$, from the most strict 0 to a
more lax 0.1. As we can see, $r$ values directly translate
to a precision/recall trade-off and is an effective
knob for different precision targets. For the proposed
FMDV-VH variant, its performance is not sensitive for
$r \geq 0.02$. 

Figure~\ref{fig:sensitivity_C} shows the effect of
varying the coverage target $m$ from 0 to 100. 
We see that the precision/recall of
our algorithms is not sensitive to $m$ in most cases,
because the test columns we sample randomly are likely
to have popular patterns matching thousands of columns.
We recommend using a large $m$ (e.g., at least 100) to ensure
confident domain inference.

Figure~\ref{fig:sensitivity_to_index_tau} shows the
impact of varying $\tau$, which is the max
number of tokens in a column for it to be indexed (Section~\ref{sec:architecture}).
As can be seen from the figure, algorithms using vertical-cuts
(FMDV-V and FMDV-VH) are insensitive to smaller
$\tau$, while algorithms without vertical-cuts
(FMDV and FMDV-H) suffer substantial recall loss with a small
$\tau=8$, which shows the benefit of using vertical-cuts.
We recommend using FMDV-VH with a small $\tau$ (e.g., 8), 
which is efficient and inexpensive to run.

Figure~\ref{fig:sensitivity_to_theta} shows the sensitivity
to $\theta$. We can see that the algorithm is not sensitive to $\theta$,
as long as it is not too small.

\iftoggle{fullversion}
{
\textbf{Pattern analysis.}
Because our offline index enumerates all possible patterns that can 
be generated from $\mathbf{T}$, it provides 
a unique opportunity to understand: (1) all common domains
in $\mathbf{T}$ independent of query-columns, defined as patterns
with high coverage and low FPR; and (2) the characteristics of all
candidate patterns generated from $\mathbf{T}$. 

For (1) we inspect the 
``head'' patterns in the index file, 
e.g. patterns matching over 10K columns and with low FPR.
This indeed reveals many
common ``domains'' of interest in this data lake, 
some examples of which are shown in 
Figure~\ref{fig:domain-example}. We note that
identifying data domains in proprietary 
formats have applications beyond 
data validation (e.g., in enterprise data search),
and is an interesting area for future research.

For (2), we show an
analysis of all patterns in $\mathbf{T}$
using the index
in Figure \ref{fig:pattern_analysis}.
Figure~\ref{fig:pattern_by_token} shows
the frequency of patterns by token-length
(each \val{<num>}, \val{<letter>}, etc.
is a token). We can see that
the patterns are fairly evenly distributed,
where patterns with 5-7 token are the most common.
Figure~\ref{fig:pattern_by_freq} shows
the frequency of all distinct candidate patterns.
We observe a power-law-like distribution -- while
the ``head'' patterns are likely useful domain patterns,
the vast majority have low coverage that are either
generalized too narrowly or are not common domains.

\begin{figure}
    \centering
    \subfigure[Pattern distr. by \#-of-tokens]{\label{fig:pattern_by_token}\includegraphics[width=0.45\columnwidth]{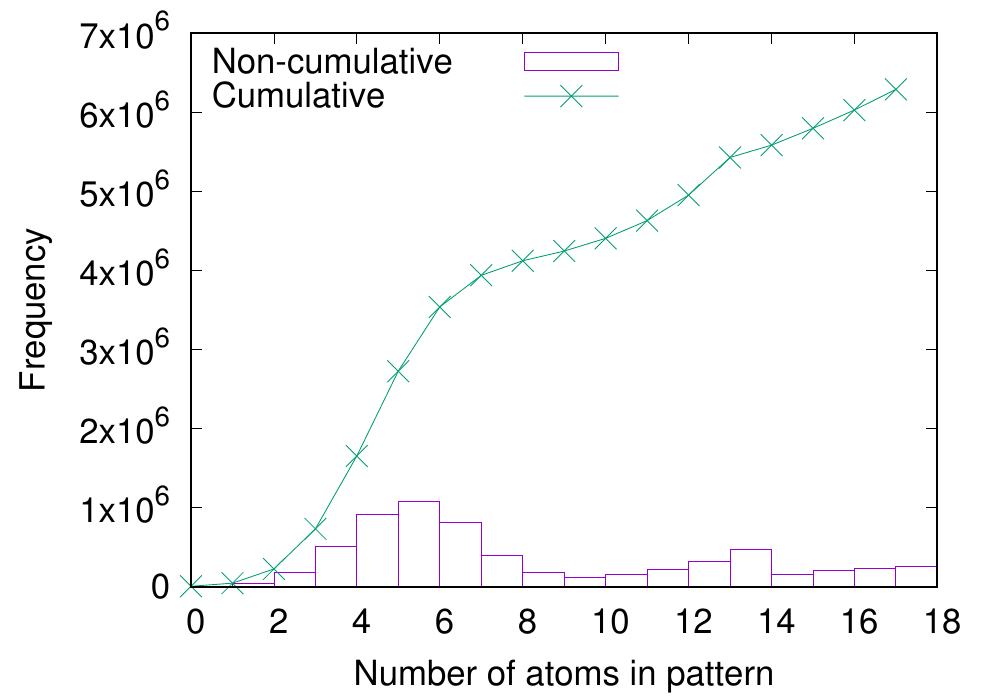}}
    \qquad
    \subfigure[Pattern distr. by frequency]{\label{fig:pattern_by_freq}\includegraphics[width=0.45\columnwidth]{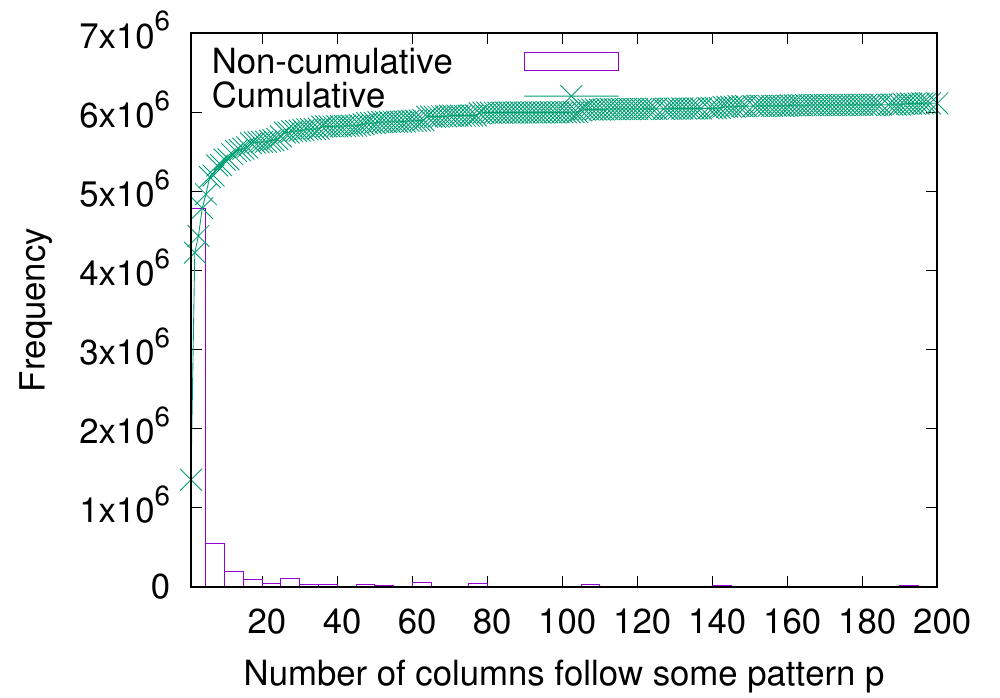}}
    
    \caption[]{Distribution of patterns in the offline index.}
    \label{fig:pattern_analysis}
\end{figure}
}

\begin{figure}
  \includegraphics[width=0.45\textwidth]{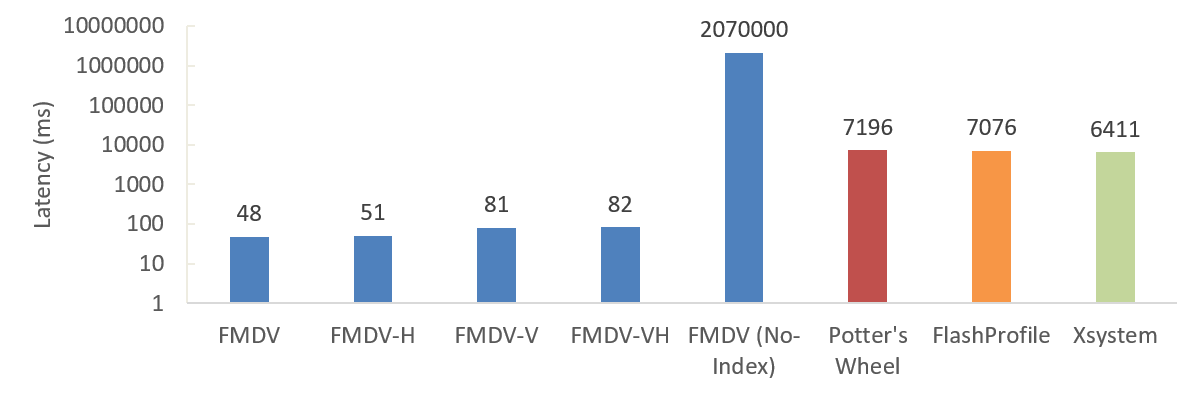}
  \caption{Latency in milliseconds (per query column).}
  \label{fig:latency}
\end{figure}

\textbf{Efficiency.}
Figure~\ref{fig:latency} shows the average latency
(in milliseconds) to process  one query column $C$ in our benchmark.
We emphasize that low latency is critical
for human-in-the-loop scenarios (e.g.,
TFDV), where users are expected to verify suggested constraints.

On the right of the figure, we can see the average latency 
for \val{PWheel}, \val{FalshProfile}, and \val{XSystem}, respectively
(we use authors' original implementations for the last two methods~\cite{FlashProfileCode, xsystem-code}).
It can be seen that these existing pattern-profiling techniques all require
on average 6-7 seconds per column. Given that tables often have tens of columns, the end-to-end latency can be slow.

In comparison, we observe that despite using a more
involved algorithm and a large corpus $\mathbf{T}$, 
all FMDV variants are two orders
of magnitude faster, 
where the most expensive FMDV-VH takes only 0.082 seconds
per column. This demonstrates the benefit of the offline indexing
step in Section~\ref{sec:architecture}, which 
distills $\mathbf{T}$ (in terabytes) down to a small index 
with summary statistics (with less than one gigabyte), 
and pushes expensive reasoning to offline, allowing 
for fast online response time.
As an additional reference point, if we do not 
use offline indexes, the ``FMDV (no-index)'' method
has to scan $\mathbf{T}$ for each query and is many orders of magnitude slower.


For the offline indexing step, we report that
the end-to-end latency of our job (on a cluster with
10 virtual-nodes) ranges from around 1 hour (with
$\tau = 8$), to around 3 hours (with $\tau = 13$).
We believe this shows that our algorithm is viable even
on small-scale clusters, despite using a large number of patterns.

\textbf{User study.}
To evaluate the human benefit of suggesting data-validation
patterns, we perform a user study by recruiting 5 
programmers (all with at least 5 years of 
programming experience),
to write data-validation patterns for 20 sampled test
columns in benchmark $\mathbf{B_E}$.  
This corresponds to the scenario of developers manually 
writing patterns without using the proposed algorithm.

\begin{table}[t]
\centering
\scriptsize
\begin{tabular}{|c | c | c | c |} 
 \hline
  Programmer & avg-time (sec) & avg-precision & avg-recall  \\  \hline
   \#1 & 145 & 0.65 & 0.638  \\  \hline 
   \#2 & 123 & 0.45 & 0.431  \\  \hline
   \#3 & 84 & 0.3 & 0.266  \\  \hline
   FMDV-VH & \textbf{0.08} & \textbf{1.0}  & \textbf{0.978} \\  \hline
\end{tabular}
\caption{User study comparing 3 developers with 
FMDV-VH on 20 test columns,  by time spent and pattern quality.}
\label{tab:user-study}
\end{table}

We report that 2 out of the 5 users fail completely
on the task (their regex are either 
ill-formed or fail to match given examples). 
Table~\ref{tab:user-study} shows the results for the
remaining 3 users. We observe that on average, they
spend 117 seconds
to write a regex for one test column (for it often
requires many trials-and-errors). This is in sharp contrast to 
0.08 seconds required by our algorithm. Programmers' average 
precision is 0.47, also substantially
lower than the algorithm's precision of 1.0 on hold-out test data.

\textbf{Case studies using Kaggle.}
In order to use publicly available data sets 
to assess the benefits of data
validation against schema-drift,
we conduct a case study using 11 tasks sampled from
Kaggle~\cite{kaggle}, where each task
has at least 2 string-valued categorical attributes.
These 11 tasks include 7 for classification: 
\val{Titanic}, \val{AirBnb}, \val{BNPParibas}, \val{RedHat}, 
\val{SFCrime}, \val{WestNile}, \val{WalmartTrips};
and 4 for regression: \val{HousePrice}, \val{HomeDepot},
\val{Caterpillar}, and \val{WalmartSales}.

To simulate
schema-drift~\cite{breck2019data, polyzotis2017data, 
schelter2018automating}, for each Kaggle task,
we keep the training data unchanged, but swap 
the position of the categorical attributes in the testing data (e.g.,
if a data set has two attributes \val{date} and \val{time} at column-position
1 and 2, respectively, after simulated 
schema-drift the two attributes will be at column-position
2 and 1, respectively). This creates a small schema-mismatch between
training/testing data that is hard to detect but can be 
common in practice like explained in~\cite{breck2019data, polyzotis2017data}. 


\begin{figure*}
  \includegraphics[width=0.7\textwidth]{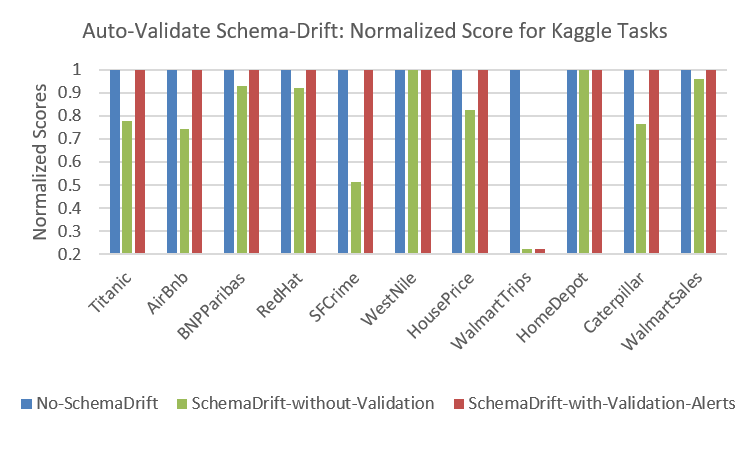}
  \caption{Impact of schema-shift on Kaggle tasks, with and without
data-validation. FMDV detects schema-drift in 
8 out of 11 cases.}
  \label{fig:kaggle}
\end{figure*}

Figure~\ref{fig:kaggle} shows the quality results with
and without validation on these Kaggle
tasks. For all tasks, we use a popular GBDT method 
called XGBoost~\cite{xgboost} with default parameters.
We report R2 for regression tasks and 
average-precision for classification tasks.

The left (blue) bar of each task, labeled as \val{No-SchemaDrift},
shows the prediction-quality scores, on the original
datasets and without schema-drift. We normalize these scores as 100\%.
The middle (green)
bar, \val{SchemaDrift-without-Validation}, shows the quality scores
with schema-drift, measured 
relative to the original quality scores.
We observe a drop up to 78\% (\val{WalmartTrips}) in normalized 
scores. 
Lastly, the right (red) bar, \val{SchemaDrift-with-Validation}
shows the quality when data-validation is used, 
which correctly detect schema-drift in 8 out of 11 tasks
(all except \val{WestNile}, \val{HomeDepot} and \val{WalmartTrips},
with no false-positives). Addressing 
such validation alerts would significantly boost the resulting quality scores.
s
While it is generally known that schema-drift
hurts ML quality~\cite{breck2019data, TFDV}, our analysis
quantifies its impact on publicly available data and
confirms the importance of data-validation.

\balance

\section{Related Works}
\label{sec:related}

\hspace{3mm}\textbf{Data Validation.}
Notable recent efforts on data validation include
Google's TensorFlow Data 
Validation (TFDV)~\cite{breck2019data, TFDV} and
Amazon's Deequ~\cite{Deequ, schelter2018automating}.
These offerings allow developers to
write declarative data quality constraints to
describe how ``normal'' data should look like, which 
are described in detail in the introduction.


\textbf{Error Detection.} There is an influential line of 
work on error-detection, including methods that focus on 
\textit{multi-column} dependencies, such as
FDs and denial constraints
(e.g.,~\cite{berti2018discovery, chiang2008discovering, chu2013,
dasu2002mining, golab2010data,  ilyas2004cords, kivinen1995approximate, yancoded}), where most techniques would require some amount
of human supervision (e.g., manually specified/verified constraints).
\sj{} is \textit{unsupervised}
and focuses on \textit{single-column} constraints, 
which naturally complements existing work.

Recent approaches propose learning-based error-detection
methods (e.g.,~\cite{heidari2019holodetect, huang2018auto, liu2020picket,
mahdavi2019raha, qahtan2019anmat, yan2020scoded, wang2019uni}), some of which are unsupervised~\cite{huang2018auto, liu2020picket, wang2019uni} and similar in spirit to \sj. 


\textbf{Pattern Profiling.}
There is also a long line of work on pattern profiling,
including pioneering methods such as
Potter's wheel~\cite{raman2001potter} that leverages
the MDL principle, and other
related techniques~\cite{ilyas2018extracting, fisher2005pads, 
fisher2008dirt, naumann2014data}.  
While both data validation 
and profiling produce patterns, data profiling
only aims to produce patterns
to ``cover'' given values in a column $C$, without
considering valid values from the same domain but not in 
$C$, and is thus prone to false-positives when used for 
data validation (for data that arrive in the future).
\nocite{chen2016metric, das2014clusterjoin}

\textbf{Other forms to validation.} 
While pattern-based validation is a natural fit for machine-generated data;
alternative forms of validation logic can be more suited for other 
types of data. For example, for natural-language data drawn from 
a fixed vocabulary (e.g., countries or airport-codes), 
dictionary-based validation learned from examples (e.g., set expansion~\cite{he2011seisa, pantel2009web, wang2007language}) 
is applicable. For complex types, semantic-type validation~\cite{hulsebos2019sherlock, 
yan2018synthesizing, zhang2019sato} is also a suitable choice.


\section{Conclusions and Future Work}

Observing the need to automate data-validation 
in production pipelines, we propose a corpus-driven approach to
inferring single-column constraints that can be used to 
auto=validate string-valued data.
Possible directions of future work include extending beyond
``machine-generated data'' to consider
natural-language-like data, and extending the same validation 
principle also to numeric data.

\clearpage



\balance
\bibliographystyle{abbrv}
{
\bibliography{auto-validate}

\begin{thebibliography}{10}

\bibitem{Deequ}
{Amazon Deequ Library for Data Validation}.
\newblock \url{https://github.com/awslabs/deequ}.

\bibitem{Glue}
{AWS Glue custom classifers}.
\newblock
  \url{https://docs.aws.amazon.com/glue/latest/dg/custom-classifier.html}.

\bibitem{azureml-pipelines}
{Azure ML: Data Pipelines}.
\newblock
  \url{https://docs.microsoft.com/en-us/azure/machine-learning/concept-ml-pipelines}.

\bibitem{purview}
{Azure Purview for data governance}.
\newblock \url{https://azure.microsoft.com/en-us/services/purview/}.

\bibitem{BingEntity}
Bing entity search.
\newblock
  \url{https://azure.microsoft.com/en-us/services/cognitive-services/bing-entity-search-api/}.

\bibitem{crawler}
{Data Crawler for NationalArchives.gov.uk}.
\newblock \url{https://github.com/alex-bogatu/DataSpiders}.

\bibitem{FlashProfileCode}
{FlashProfile package}.
\newblock
  \url{https://www.nuget.org/packages/Microsoft.ProgramSynthesis.Extraction.Text/}.

\bibitem{TFDV}
{Google TensorFlow Data Validation}.
\newblock \url{https://www.tensorflow.org/tfx/guide/tfdv}.

\bibitem{grok}
{Grok Patterns}.
\newblock
  \url{https://github.com/elastic/elasticsearch/blob/master/libs/grok/src/main/resources/patterns/grok-patterns}.

\bibitem{Informatica}
{Informatica Rev}.
\newblock \url{https://www.informatica.com/products/data-quality/rev.html}.

\bibitem{kaggle}
Kaggle.
\newblock \url{https://www.kaggle.com/}.

\bibitem{power-bi-flow}
{Power BI: Data Flow}.
\newblock
  \url{https://docs.microsoft.com/en-us/power-bi/transform-model/service-dataflows-create-use}.

\bibitem{ssis-profiling}
{SSIS: Data Profiling}.
\newblock
  \url{https://docs.microsoft.com/en-us/sql/integration-services/control-flow/data-profiling-task?view=sql-server-ver15}.

\bibitem{tableau-flow}
{Tableau: Flow}.
\newblock
  \url{https://help.tableau.com/current/prep/en-us/prep_build_flow.htm}.

\bibitem{amazon-workflow}
{Tableau: Flow}.
\newblock
  https://aws.amazon.com/blogs/big-data/simplify-data-pipelines-with-aws-glue-automatic-code-generation-and-workflows/.

\bibitem{xgboost}
{XGBoost}.
\newblock \url{https://xgboost.readthedocs.io/en/latest/}.

\bibitem{xsystem-code}
{XSystem Code}.
\newblock
  \url{https://bitbucket.org/andrewiilyas/xsystem-old/src/outlier-detection/}.

\bibitem{agresti1992survey}
A.~Agresti et~al.
\newblock A survey of exact inference for contingency tables.
\newblock {\em Statistical science}, 7(1):131--153, 1992.

\bibitem{berti2018discovery}
L.~Berti-Equille, H.~Harmouch, F.~Naumann, N.~Novelli, and
  S.~Thirumuruganathan.
\newblock Discovery of genuine functional dependencies from relational data
  with missing values.
\newblock {\em VLDB}, 2018.

\bibitem{bogatu2020dataset}
A.~Bogatu, A.~A. Fernandes, N.~W. Paton, and N.~Konstantinou.
\newblock Dataset discovery in data lakes.
\newblock In {\em 2020 IEEE 36th International Conference on Data Engineering
  (ICDE)}, pages 709--720. IEEE, 2020.

\bibitem{breck2019data}
E.~Breck, N.~Polyzotis, S.~Roy, S.~Whang, and M.~Zinkevich.
\newblock Data validation for machine learning.
\newblock In {\em Conference on Systems and Machine Learning (SysML).
  https://www. sysml. cc/doc/2019/167. pdf}, 2019.

\bibitem{carrillo1988multiple}
H.~Carrillo and D.~Lipman.
\newblock The multiple sequence alignment problem in biology.
\newblock {\em SIAM journal on applied mathematics}, 48(5):1073--1082, 1988.

\bibitem{chaiken2008scope}
R.~Chaiken, B.~Jenkins, P.-{\AA}. Larson, B.~Ramsey, D.~Shakib, S.~Weaver, and
  J.~Zhou.
\newblock Scope: easy and efficient parallel processing of massive data sets.
\newblock {\em Proceedings of the VLDB Endowment}, 1(2):1265--1276, 2008.

\bibitem{chen2016metric}
G.~Chen, K.~Yang, L.~Chen, Y.~Gao, B.~Zheng, and C.~Chen.
\newblock Metric similarity joins using mapreduce.
\newblock {\em IEEE Transactions on Knowledge and Data Engineering},
  29(3):656--669, 2016.

\bibitem{chiang2008discovering}
F.~Chiang and R.~J. Miller.
\newblock Discovering data quality rules.
\newblock {\em VLDB}, 1(1), 2008.

\bibitem{chu2013}
X.~Chu, I.~F. Ilyas, and P.~Papotti.
\newblock Discovering denial constraints.
\newblock {\em VLDB}, 6(13), 2013.

\bibitem{cormen2009introduction}
T.~H. Cormen, C.~E. Leiserson, R.~L. Rivest, and C.~Stein.
\newblock {\em Introduction to algorithms}.
\newblock MIT press, 2009.

\bibitem{das2014clusterjoin}
A.~Das~Sarma, Y.~He, and S.~Chaudhuri.
\newblock Clusterjoin: A similarity joins framework using map-reduce.
\newblock {\em Proceedings of the VLDB Endowment}, 7(12):1059--1070, 2014.

\bibitem{dasu2002mining}
T.~Dasu, T.~Johnson, S.~Muthukrishnan, and V.~Shkapenyuk.
\newblock Mining database structure; or, how to build a data quality browser.
\newblock In {\em SIGMOD}, 2002.

\bibitem{dayal2009data}
U.~Dayal, M.~Castellanos, A.~Simitsis, and K.~Wilkinson.
\newblock Data integration flows for business intelligence.
\newblock In {\em Proceedings of the 12th International Conference on Extending
  Database Technology: Advances in Database Technology}, pages 1--11, 2009.

\bibitem{fisher2005pads}
K.~Fisher and R.~Gruber.
\newblock Pads: a domain-specific language for processing ad hoc data.
\newblock {\em ACM Sigplan Notices}, 40(6):295--304, 2005.

\bibitem{fisher2008dirt}
K.~Fisher, D.~Walker, K.~Q. Zhu, and P.~White.
\newblock From dirt to shovels: fully automatic tool generation from ad hoc
  data.
\newblock {\em ACM SIGPLAN Notices}, 43(1):421--434, 2008.

\bibitem{golab2010data}
L.~Golab, H.~Karloff, F.~Korn, and D.~Srivastava.
\newblock Data auditor: Exploring data quality and semantics using pattern
  tableaux.
\newblock {\em VLDB}, 3(1-2), 2010.

\bibitem{auto-tag-tr}
Y.~He, J.~Song, Y.~Wang, S.~Chaudhuri, V.~Anil, B.~Lassiter, Y.~Goland, and
  G.~Malhotra.
\newblock Auto-tag: Tagging-data-by-example in data lakes using pre-training
  and inferred domain patterns.

\bibitem{he2011seisa}
Y.~He and D.~Xin.
\newblock Seisa: set expansion by iterative similarity aggregation.
\newblock In {\em Proceedings of the 20th international conference on World
  wide web}, pages 427--436, 2011.

\bibitem{heidari2019holodetect}
A.~Heidari, J.~McGrath, I.~F. Ilyas, and T.~Rekatsinas.
\newblock Holodetect: Few-shot learning for error detection.
\newblock In {\em Proceedings of the 2019 International Conference on
  Management of Data}, pages 829--846, 2019.

\bibitem{huang2018auto}
Z.~Huang and Y.~He.
\newblock {Auto-Detect: Data-Driven Error Detection in Tables}.
\newblock In {\em SIGMOD}, 2018.

\bibitem{hulsebos2019sherlock}
M.~Hulsebos, K.~Hu, M.~Bakker, E.~Zgraggen, A.~Satyanarayan, T.~Kraska,
  {\c{C}}.~Demiralp, and C.~Hidalgo.
\newblock Sherlock: A deep learning approach to semantic data type detection.
\newblock In {\em Proceedings of the 25th ACM SIGKDD International Conference
  on Knowledge Discovery \& Data Mining}, pages 1500--1508, 2019.

\bibitem{hynes2017data}
N.~Hynes, D.~Sculley, and M.~Terry.
\newblock The data linter: Lightweight, automated sanity checking for ml data
  sets.
\newblock In {\em NIPS MLSys Workshop}, 2017.

\bibitem{ilyas2018extracting}
A.~Ilyas, J.~M. da~Trindade, R.~C. Fernandez, and S.~Madden.
\newblock Extracting syntactical patterns from databases.
\newblock In {\em 2018 IEEE 34th International Conference on Data Engineering
  (ICDE)}, pages 41--52. IEEE, 2018.

\bibitem{ilyas2004cords}
I.~F. Ilyas, V.~Markl, P.~Haas, P.~Brown, and A.~Aboulnaga.
\newblock Cords: automatic discovery of correlations and soft functional
  dependencies.
\newblock In {\em SIGMOD}, 2004.

\bibitem{just2001computational}
W.~Just.
\newblock Computational complexity of multiple sequence alignment with
  sp-score.
\newblock {\em Journal of computational biology}, 8(6):615--623, 2001.

\bibitem{kanji2006}
G.~K. Kanji.
\newblock {\em 100 statistical tests}.
\newblock Sage, 2006.

\bibitem{karp1985fast}
R.~M. Karp and A.~Wigderson.
\newblock A fast parallel algorithm for the maximal independent set problem.
\newblock {\em Journal of the ACM (JACM)}, 32(4):762--773, 1985.

\bibitem{kivinen1995approximate}
J.~Kivinen and H.~Mannila.
\newblock Approximate inference of functional dependencies from relations.
\newblock {\em Theoretical Computer Science}, 149(1), 1995.

\bibitem{liu2020picket}
Z.~Liu, Z.~Zhou, and T.~Rekatsinas.
\newblock Picket: Self-supervised data diagnostics for ml pipelines.
\newblock {\em arXiv preprint arXiv:2006.04730}, 2020.

\bibitem{mahdavi2019raha}
M.~Mahdavi, Z.~Abedjan, R.~Castro~Fernandez, S.~Madden, M.~Ouzzani,
  M.~Stonebraker, and N.~Tang.
\newblock Raha: A configuration-free error detection system.
\newblock In {\em Proceedings of the 2019 International Conference on
  Management of Data}, pages 865--882, 2019.

\bibitem{naumann2014data}
F.~Naumann.
\newblock Data profiling revisited.
\newblock {\em ACM SIGMOD Record}, 42(4):40--49, 2014.

\bibitem{padhi2018flashprofile}
S.~Padhi, P.~Jain, D.~Perelman, O.~Polozov, S.~Gulwani, and T.~Millstein.
\newblock Flashprofile: a framework for synthesizing data profiles.
\newblock {\em Proceedings of the ACM on Programming Languages},
  2(OOPSLA):1--28, 2018.

\bibitem{pantel2009web}
P.~Pantel, E.~Crestan, A.~Borkovsky, A.-M. Popescu, and V.~Vyas.
\newblock Web-scale distributional similarity and entity set expansion.
\newblock In {\em Proceedings of the 2009 Conference on Empirical Methods in
  Natural Language Processing}, pages 938--947, 2009.

\bibitem{papenbrock2016hybrid}
T.~Papenbrock and F.~Naumann.
\newblock A hybrid approach to functional dependency discovery.
\newblock In {\em Proceedings of the 2016 International Conference on
  Management of Data}, pages 821--833, 2016.

\bibitem{patel2019big}
H.~Patel, A.~Jindal, and C.~Szyperski.
\newblock Big data processing at microsoft: Hyper scale, massive complexity,
  and minimal cost.
\newblock In {\em Proceedings of the ACM Symposium on Cloud Computing}, pages
  490--490, 2019.

\bibitem{polyzotis2017data}
N.~Polyzotis, S.~Roy, S.~E. Whang, and M.~Zinkevich.
\newblock Data management challenges in production machine learning.
\newblock In {\em Proceedings of the 2017 ACM International Conference on
  Management of Data}, pages 1723--1726, 2017.

\bibitem{qahtan2019anmat}
A.~Qahtan, N.~Tang, M.~Ouzzani, Y.~Cao, and M.~Stonebraker.
\newblock Anmat: automatic knowledge discovery and error detection through
  pattern functional dependencies.
\newblock In {\em Proceedings of the 2019 International Conference on
  Management of Data}, pages 1977--1980, 2019.

\bibitem{rahm2001survey}
E.~Rahm and P.~A. Bernstein.
\newblock A survey of approaches to automatic schema matching.
\newblock {\em the VLDB Journal}, 10(4):334--350, 2001.

\bibitem{rahm2000data}
E.~Rahm and H.~H. Do.
\newblock Data cleaning: Problems and current approaches.
\newblock {\em IEEE Data Eng. Bull.}, 23(4):3--13, 2000.

\bibitem{raman2001potter}
V.~Raman and J.~M. Hellerstein.
\newblock Potter's wheel: An interactive data cleaning system.
\newblock In {\em VLDB}, volume~1, 2001.

\bibitem{schelter2019unit}
S.~Schelter, F.~Biessmann, D.~Lange, T.~Rukat, P.~Schmidt, S.~Seufert,
  P.~Brunelle, and A.~Taptunov.
\newblock Unit testing data with deequ.
\newblock In {\em Proceedings of the 2019 International Conference on
  Management of Data}, pages 1993--1996, 2019.

\bibitem{schelter2019differential}
S.~Schelter, S.~Grafberger, P.~Schmidt, T.~Rukat, M.~Kiessling, A.~Taptunov,
  F.~Biessmann, and D.~Lange.
\newblock Differential data quality verification on partitioned data.
\newblock In {\em 2019 IEEE 35th International Conference on Data Engineering
  (ICDE)}, pages 1940--1945. IEEE, 2019.

\bibitem{schelter2018automating}
S.~Schelter, D.~Lange, P.~Schmidt, M.~Celikel, F.~Biessmann, and A.~Grafberger.
\newblock Automating large-scale data quality verification.
\newblock {\em Proceedings of the VLDB Endowment}, 11(12):1781--1794, 2018.

\bibitem{stonebraker2018data}
M.~Stonebraker and I.~F. Ilyas.
\newblock Data integration: The current status and the way forward.
\newblock {\em IEEE Data Eng. Bull.}, 41(2):3--9, 2018.

\bibitem{swami2020data}
A.~Swami, S.~Vasudevan, and J.~Huyn.
\newblock Data sentinel: A declarative production-scale data validation
  platform.
\newblock In {\em 2020 IEEE 36th International Conference on Data Engineering
  (ICDE)}, pages 1579--1590. IEEE, 2020.

\bibitem{vassiliadis2009near}
P.~Vassiliadis and A.~Simitsis.
\newblock Near real time etl.
\newblock In {\em New trends in data warehousing and data analysis}, pages
  1--31. Springer, 2009.

\bibitem{wang2019uni}
P.~Wang and Y.~He.
\newblock Uni-detect: A unified approach to automated error detection in
  tables.
\newblock In {\em Proceedings of the 2019 International Conference on
  Management of Data}, pages 811--828, 2019.

\bibitem{wang2007language}
R.~C. Wang and W.~W. Cohen.
\newblock Language-independent set expansion of named entities using the web.
\newblock In {\em Seventh IEEE international conference on data mining (ICDM
  2007)}, pages 342--350. IEEE, 2007.

\bibitem{yan2018synthesizing}
C.~Yan and Y.~He.
\newblock {Auto-Type: Synthesizing type-detection logic for rich semantic data
  types using open-source code}.
\newblock In {\em Proceedings of the 2018 International Conference on
  Management of Data}. ACM, 2018.

\bibitem{yancoded}
J.~N. Yan, O.~Schulte, J.~Wang, and R.~Cheng.
\newblock Coded: Column-oriented data error detection with statistical
  constraints.

\bibitem{yan2020scoded}
J.~N. Yan, O.~Schulte, M.~Zhang, J.~Wang, and R.~Cheng.
\newblock Scoded: Statistical constraint oriented data error detection.
\newblock In {\em Proceedings of the 2020 ACM SIGMOD International Conference
  on Management of Data}, pages 845--860, 2020.

\bibitem{zhang2019sato}
D.~Zhang, Y.~Suhara, J.~Li, M.~Hulsebos, {\c{C}}.~Demiralp, and W.-C. Tan.
\newblock Sato: Contextual semantic type detection in tables.
\newblock {\em arXiv preprint arXiv:1911.06311}, 2019.

\bibitem{zhou2012scope}
J.~Zhou, N.~Bruno, M.-C. Wu, P.-A. Larson, R.~Chaiken, and D.~Shakib.
\newblock Scope: parallel databases meet mapreduce.
\newblock {\em The VLDB Journal}, 21(5):611--636, 2012.

\end{thebibliography}
}


\end{document}